\documentclass[review,hidelinks,onefignum,onetabnum]{siamart220329}

\usepackage{lipsum}
\usepackage{amsfonts,amssymb}
\usepackage{graphicx}
\usepackage{epstopdf}
\usepackage{algorithmic}
\ifpdf
  \DeclareGraphicsExtensions{.eps,.pdf,.png,.jpg}
\else
  \DeclareGraphicsExtensions{.eps}
\fi

\newcommand{\RR}{\mathbb{R}}

\newcommand{\ve}{\varepsilon}

\newcommand{\bz}{\mathbf{z}}
\newcommand{\bx}{\mathbf{x}}
\newcommand{\by}{\mathbf{y}}
\newcommand{\bk}{\mathbf{k}}
\newcommand{\bq}{\mathbf{q}}
\newcommand{\bp}{\mathbf{p}}
\newcommand{\bK}{\mathbf{K}}

\newcommand{\bnu}{\bm{\nu}}
\newcommand{\bmu}{\bm{\mu}}
\newcommand{\expec}{\mathbb{E}}

\newcommand*{\im}{\mathop{}\!\mathrm{i}}
\newcommand*{\e}{\mathop{}\!\mathrm{e}}

\newcommand{\eqdef }{\overset{\mbox{\tiny{def}}}{=}}

\newcommand{\rd}{\mathrm{d}}

\newsiamremark{remark}{Remark}
\newsiamremark{hypothesis}{Hypothesis}
\crefname{hypothesis}{Hypothesis}{Hypotheses}
\newsiamthm{claim}{Claim}

\headers{Radiative transport in a periodic structure with band crossings}{K.~Qi, L.~Wang, and A.~B.~Watson}

\title{Radiative transport in a periodic structure with band crossings \thanks{Submitted to the editors DATE.
\funding{LW is partially supported by NSF grant DMS-1846854. AW's research was supported in part by NSF grants DMREF 1922165 and DMS 2406981.}}}

\author{Kunlun Qi \and Li Wang \and Alexander B. Watson \thanks{School of Mathematics, University of Minnesota--Twin Cities, Minneapolis, MN 55455 USA. (\email{kqi@umn.edu}), (\email{liwang@umn.edu}), (\email{abwatson@umn.edu}).}}

\usepackage{amsopn}
\usepackage{bm}

\ifpdf
\hypersetup{
  pdftitle={Radiative transport in a periodic structure with band crossings},
  pdfauthor={K.~Qi L.~Wang and A.~B.~ Watson}
}
\fi

\begin{document}
\nolinenumbers

\maketitle

% REQUIRED
\begin{abstract}
We use the Wigner transformation and asymptotic analysis to systematically derive the semi-classical model for the Schr\"{o}dinger equation in arbitrary spatial dimensions, with any periodic structure. Our particular emphasis lies in addressing the \textit{diabatic} effect, i.e., the impact of Bloch band crossings. We consider both deterministic and random scenarios. In the former case, we derive a coupled Liouville system, revealing lower-order interactions among different Bloch bands. In the latter case, a coupled system of radiative transport equations emerges, with the scattering cross-section induced by the random inhomogeneities. As a specific application, we deduce the effective dynamics of a wave packet in graphene with randomness.
\end{abstract}

% REQUIRED

% REQUIRED

\section{Introduction}
\label{sec:intro}
The Schr\"{o}dinger equation, serving as a model that describes the evolution of the quantum state of a physical system, such as an electron, over time, has found widespread application in solid-state physics. The equation is given by
\begin{equation}\label{s}
 \im  \partial_t \phi(t,\bx) + \frac{1}{2} \Delta \phi(t,\bx) - V\left( \bx \right) \phi (t,\bx) =\  0, \quad t \in \RR_{+}, \quad \bx \in \RR^d\,,
\end{equation}
where $\phi(t,\bx)$ represents the complex-valued wave function.
When the material displays a certain lattice structure, characterized by the periodicity in $V$ (i.e., $V(\bx + \bnu) = V(\bx)$ for any $\bx$ in $\RR^d$ and $\bnu$ being the lattice vector), Floquet–Bloch theory offers insight into the energy band structures of the material (elaborated in Section~\ref{sec21}). The interaction of these bands determines the {\it adiabatic} and {\it diabatic} behavior of the quantum system, depending on whether the bands intersect.

The latter scenario is particularly evident in structures with honeycomb lattice symmetry, such as graphene, a two-dimensional material composed of a single layer of carbon atoms \cite{2005NovoselovGeimMorozovJiangKatsnelsonGrigorievaDubonosFirsov}. Indeed, the energy dispersion surfaces of graphene reveal conical singularities at the intersections of the valence and conduction bands \cite{FW2012JAMS, FW2014CMP, 2018BerkolaikoComech}. These singularities, termed as {\it Dirac points}, form cone-like shapes centered at the vertices of the Brillouin zone, playing a pivotal role in its remarkable electronic and mechanical properties \cite{NGPNG2009}. Graphene has attracted renewed attention in recent years since the observation that magic angle twisted bilayer graphene, i.e., two layers of graphene, stacked with a relative twist $\approx 1^\circ$, displays signs of unconventional superconductivity \cite{Bistritzer2011, Cao2018, Cao2018a}. 
%\aw{I revised the last sentence to clarify that TBG superconducts. Graphene on its own doesn't.}
%Notably, graphene exhibits superconductivity at the magic angle \cite{Bistritzer2011, Cao2018, Cao2018a} and various other phenomena extensively reviewed in .

When the initial data of \eqref{s} is concentrated at the Dirac points, Fefferman and Weinstein have demonstrated that the solution to \eqref{s} can be expressed as the superposition of Floquet-Bloch states with modulated amplitudes. These amplitudes adhere to the Dirac equations \cite{FW2014CMP}. Consequently, an intriguing question arises: In the presence of material perturbations, such as impurities, what impact will it have on the Dirac equations?

To tackle this problem, rather than directly modifying the Dirac equations, we opt for a different approach by revisiting the Schr\"{o}dinger equation and introducing a random potential to model the perturbation. Instead of following the derivation in \cite{FW2014CMP}, we consider a different scenario by incorporating a semiclassical scaling into the Schr\"{o}dinger equation. This scaling facilitates the treatment of randomness. Specifically, under this scaling, the equation \eqref{s} rewrites into:
\begin{equation} \label{s5}
\im \ve \partial_t \phi_{\ve} + \frac{\ve^2}{2} \Delta_\bx \phi_{\ve} - V\left( \frac{\bx}{\ve} \right) \phi_{\ve} - \sqrt{\ve} N\left( \frac{\bx}{\ve} \right)  \phi_{\ve} =0\,,
\end{equation}
where $\varepsilon$ is the rescaled Planck constant, and $N$ denotes the random potential. The fundamental question now becomes: What is the semiclassical limit (i.e., $\varepsilon \downarrow 0$) of \eqref{s5} in the presence of band crossings?

%\aw{I suggest removing this paragraph here. I think it works better where I copied it below and added some more detail about originality.} We have shown that, in the absence of randomness, the semiclassical limit of \eqref{s5} in the vicinity of the crossing point is a system of Liouville equations, reminiscent of the Dirac system of \cite{FW2014CMP}, with an additional relaxation-type source term, see \eqref{eq:W_Dirac}. This term vanishes at the crossing point, thereby reducing the system to the homogeneous case. In the presence of randomness, an additional integral term emerges, reminiscent of the collision term in the radiative transport equation, see \eqref{eq:final_system}. This term describes the coupling between distinct wave vectors with the same energy.

The primary tool we utilize here is the Wigner transform, which lifts the wave function from physical space to phase space by adding the dependence on quasi-momenta, and therefore bears a close analogy to the classical mechanics \cite{Wigner1932}. When both $V$ and $N$ are absent in \eqref{s5}, the semiclassical limit yields the Liouville equation. This result is rigorously established by G\'erard in \cite{Gerard1991}, and subsequently confirmed by Lions and Paul in \cite{LP1993}, and Markowich and Mauser in \cite{markowich1993classical}.
When a periodic function $V$ exists, Markowich, Mauser, and Poupaud in \cite{MMP1994} have rigorously demonstrated that, following the approach outlined in \cite{LP1993}, provided the bands are well-separated, the semiclassical limit remains the Liouville equation traversing each band. This result has been significantly expanded upon by Bechouche, Mauser, and Poupaud in \cite{BMP2001}, including both periodic and non-periodic potentials, as well as the homogenization limit. There, a variation of the Wigner transform known as the Wigner-Bloch series has been introduced to handle the density matrices associated with two energy bands. All these works focus on the {\it adabatic} dynamics. 

To account for the {\it diabatic} dynamics, the well-known Landau-Zener formula quantifies the probability of a transition occurring between two energy levels \cite{Landau1932, Zener1932}. It has since then been extensively studied in the context of the surface hopping method, beyond the Born-Oppenheimer approximation. We refer the readers for analytical works in \cite{Hagedorn1994, Lasser2003, Fermanian-Kammerer2004, Lasser2005, Lasser2007, 2013Fermanian-KammererGerardLasser, CJL2013KRM, 2018WatsonWeinstein} and computational studies in \cite{TP1971, ST1998, Drukker1999}. We mention in particular Chai, Jin, and Li in \cite{CJL2013KRM} who derived the semiclassical limit in the form of a coupled Liouville equation near band crossing points in one dimension.

In the presence of randomness, a radiative transport-type equation is anticipated in the semiclassical limit. This was initially identified by Spohn in \cite{Spohn1977}, where he rigorously derived such a limit for time-dependent Gaussian random impurities. However, his result is limited to short times. Subsequently, Ho, Landau, and Wilkins refined Spohn's work by considering higher-order corrections in \cite{HLW1993}. Erd\"os and Yau further extended these findings to include general initial conditions and removed the restriction on the smallness of time in \cite{EY2000CPAM}. Extensions of these findings to more general types of waves in random media, such as those described by hyperbolic systems, can be found in the work of Ryzhik, Papanicolaou, and Keller \cite{RPK1996}. Further extensions, including the incorporation of periodic structures, are addressed by Bal, Fannjiang, Papanicolaou, and Ryzhik in \cite{BFPR1999JSP}, while the addition of nonlinear terms is investigated by Fannjiang, Jin, and Papanicolaou \cite{FJP2003}. A rigorous derivation of the radiative transport limit of the Schr\"{o}dinger equation is established by Bal, Papanicolaou, and Ryzhik in \cite{BPR2002}, assuming time-dependent randomness and utilizing the concept of martingales. More recently, the inclusion of randomness in the Dirac equation has been explored by Bal, Gu, and Pinaud \cite{BGP2018}.

Building upon previous discoveries, in this work, we combine the techniques of the Wigner transform for handling semiclassical scaling and randomness with Floquet-Bloch theory to address periodicity. Our goal is to derive a coupled system describing energy propagation in the semiclassical limit, accounting both for the coupling between bands at band crossings, and for coupling between distinct wave-vectors induced by randomness. 

%We have shown 
%\aw{Paragraph above moved here:} 
The results of this paper are as follows. In the absence of randomness, the semiclassical limit of \eqref{s5} in the vicinity of the crossing point is a system of Liouville equations, reminiscent of the Dirac system of \cite{FW2014CMP}, with an additional relaxation-type source term, see \eqref{eq:W_Dirac}. This term vanishes at the crossing point, thereby reducing the system to the homogeneous case. Here, we generalize the results of Chai, Jin, and Li \cite{CJL2013KRM} to higher dimensions. In the presence of randomness, an additional integral term emerges, reminiscent of the collision term in the radiative transport equation, see \eqref{eq:final_system}. This term describes the coupling between distinct wave vectors with the same energy. 
Here, we generalize the results of Bal, Fannjiang, Papanicolaou, and Ryzhik \cite{BFPR1999JSP}, whose methods do not immediately extend to the case where bands have non-constant multiplicities because of band crossings, as occurs in graphene at ``Dirac points". To our knowledge, the models \eqref{sigma_tilde_random}, \eqref{eq:reduced_graphene_0}, and \eqref{eq:reduced_graphene_minus1}-\eqref{eq:reduced_graphene_minus2} of the dynamics of electrons in materials with band crossings in the presence of randomness, modeled by \eqref{s5}, are original to the present work. The main difficulty in deriving these models is to address the subtle effect where band crossings not only affect the dynamics for quasimomenta precisely at the crossing point, but also within an $\varepsilon$-dependent neighborhood of the crossing point. Within this neighborhood, the effects of the band crossing are of order $1$ for quasimomenta. The size of this neighborhood is $\varepsilon$-dependent and varies according to the precise form of the bands near the crossing point; see the discussions in Sections \ref{sec:reduction}, \ref{sec:with_randomness}, \ref{sec:application}, and Appendix \ref{appendix-C}.

The rest of the paper is organized as follows. In the next section, we review some preliminary results concerning the Schr\"{o}dinger equation with periodic potential, its associated Bloch theory, and the Wigner transform. Section \ref{sec:without_randomness} focuses on scenarios without randomness, where we derive a coupled Liouville system featuring a relaxation-type source term to account for band crossing effects. In Section \ref{sec:with_randomness}, we extend our analysis by introducing random perturbations, leading to the derivation of a coupled radiative transport system.
The results are then applied to the honeycomb structure in Section \ref{sec:application}, where we elucidate the effective dynamics of wave packets in graphene with randomness.
Finally, the paper is concluded in Section \ref{sec:conclusions}.

%%%%%%%%%%%%%%%%%%%%%%%%%%%%%%%%%%%%%%%%%%%%%%%%%%%%%%

\section{Preliminaries}
\label{sec:preliminary}

\subsection{The Floquet-Bloch theory} \label{sec21}
% Consider the dimensionless form of the Schr\"{o}dinger equation given by: 
% \begin{equation}\label{s}
%     \left\{
%     \begin{aligned}
%     & \im  \frac{\partial \phi(t,\bx)}{\partial t} + \frac{1}{2} \Delta \phi(t,\bx) - V\left( \bx \right) \phi (t,\bx) =\  0, \quad t \in \RR_{+}, \quad \bx \in \RR^d, \\
%     & \phi (t=0,\bx) =\ \phi^0 (\bx)\,.
%     \end{aligned}
%     \right.
% \end{equation}
% Here, the initial condition $\phi^0(\bx)$ is assumed to be in $L^2(\RR^d)$, i.e., $\|\phi^0(\cdot)\|_{L^2(\RR^d)} \leq C$. Then owing to the mass conservation, the solution $\phi(t,\bx)$ remains in $L^2(\RR^d)$:
% \begin{equation*}
%     \| \phi(t,\cdot) \|_{L^2(\RR^d)} = \|\phi^0(\cdot)\|_{L^2(\RR^d)} \leq C, \quad \forall\, t \geq 0.
% \end{equation*}

Let $\{ \mathbf{e}_1,..., \mathbf{e}_d\}$ be a basis in $\RR^d$, we define a periodic lattice $\Lambda$ as: 
\begin{equation*}
    \Lambda \eqdef \big\{ \sum_{j=1}^d m_j \mathbf{e}_j \, \big| \, m_j \in \mathbb{Z}, \, j=1,...,d \big\} \,.
    %= \mathbb{Z} \mathbf{e}_1 \oplus  ... \oplus \mathbb{Z} \mathbf{e}_d
\end{equation*}
Then its dual lattice, denoted by $\Lambda^*$, is defined as: 
\begin{equation*}
    \Lambda^* \eqdef \big\{ \sum_{l=1}^d m_l \mathbf{e}^l \, \big| \, m_l \in \mathbb{Z}, \, l=1,...,d \big\} \,,
    %= \mathbb{Z} \mathbf{e}^1 \oplus  ... \oplus \mathbb{Z} \mathbf{e}^d
\end{equation*}
where $\mathbf{e}^l$ is the dual basis of $\mathbf{e}_j$ in the sense that:
\begin{equation*}
    (\mathbf{e}_j \cdot \mathbf{e}^l ) = 2 \pi \delta_{jl}.
\end{equation*}

Additionally, we define the fundamental cell of $\Lambda$ as $\mathcal{C}$:
\begin{equation*}
    \mathcal{C} \eqdef \big\{ \sum_{j=1}^d \theta_j \mathbf{e}_j \, \big| \, 0 \leq \theta_j < 1, \, j=1,...,d \big\},
\end{equation*}
and the Brillouin zone as $\mathcal{B}$:
\begin{equation*}
    %\mathcal{B} \eqdef \big\{ \mathbf{k} \in \RR^d \, \big| \, \mathbf{k} \, \text {is closer to}\, \bm{\bmu} = \mathbf{0}\, \text { than any other point}\, \bm{\bmu} \in \Lambda^* \big\}\,. 
    \mathcal{B} \eqdef \big\{ \sum_{l = 1}^d \theta_l \bm{\e}^l \, \big| \, 0 \leq \theta_l < 1, \, l = 1,...,d \big\}\,. %\mathbf{k} \in \RR^d \, \big| \, \mathbf{k} \, \text {is closer to}\, \bm{\bmu} = \mathbf{0}\, \text { than any other point}\, \bm{\bmu} \in \Lambda^* \big\}\,. 
\end{equation*}

Throughout this paper, we will consider the real-valued potential function $V(\bx)$ with $\Lambda$ periodicity, i.e.,
\begin{equation*}
    V(\bx + \bnu) = V(\bx), \quad \forall \, \bx \in \RR^d, \quad \bnu \in \Lambda. 
\end{equation*}
Denote 
\begin{equation*}
    H_{V} \eqdef -\frac{1}{2} \Delta_\bx + V(\bx).
\end{equation*}
For $\bk \in \mathbb{R}^d$, we can consider the \textit{Bloch} eigenvalue problem
%Since $H_V$ is self adjoint w.r.t. $L^2(\mathcal{C})$, it has a compact resolvent. Consequently, it features a countable family of eigenfunctions $\Psi_m(\bx, \bk)$ for a given $\bk \in \RR^d$ that satisfies
\begin{equation}\label{FB}
    \left\{
    \begin{aligned}
    H_{V} \Psi(\bx,\bk) = &\ E(\bk) \Psi(\bx,\bk)\, , \\[8pt]
    \Psi(\bx + \bnu,\bk) =&\ \e^{\im \bk \cdot \bnu } \Psi(\bx,\bk), \quad \forall \, \bnu \in \Lambda \, ,\\[8pt]
    \frac{\partial \Psi(\bx + \bnu,\bk)}{\partial x_j} =&\ \e^{\im \bk \cdot \bnu } \frac{\partial \Psi(\bx,\bk)}{\partial x_j}, \quad \forall \, \bnu \in \Lambda, \quad j = 1 , \cdots, d\,.
    \end{aligned}
    \right.
\end{equation}
We now summarize the essential properties of this eigenvalue problem. For more details, see \cite{BLG1978, OK1964, Wilcox1978}.

%which are necessary in the remainder of the paper.
 %In the following, we list a couple of essential properties of $\Psi_m(\bx,\bk)$ that will be utilized throughout the paper.
 \begin{itemize}
     \item[i)] For each $\bk$, the eigenvalue problem \eqref{FB} is self-adjoint with compact resolvent. Hence, it has a real and discrete spectrum which can be ordered with the multiplicity
     \begin{equation*}
    E_1(\bk) \leq E_2(\bk) \leq ... \leq E_m(\bk) \leq ... \quad \text{with} \quad E_m(\bk) \rightarrow \infty \quad \text{as} \ m \rightarrow \infty\,.
     \end{equation*}
     The associated Bloch eigenfunctions $\Psi_{m}(\bx,\bk)$ form a complete orthonormal basis in $L^2(\mathcal C)$ with
\begin{equation*} %\label{orth1}
    \frac{1}{|\mathcal{C}|} \int_{\mathcal{C}} \Psi_{m}(\bx,\bk) \,\overline{\Psi_{j}(\bx,\bk)} \,\rd \bx = \delta_{mj}\,.
\end{equation*}
Here $\delta_{mj}$ is the Kronecker delta function.
     \item[ii)] The eigenvalue problem \eqref{FB} is invariant under replacement of $\bk$ by $\bk + \bmu$ for $\bmu \in \Lambda^*$. Therefore it suffices to consider $\bk \in \mathcal{B}$, and that the eigenvalue band functions $E_{m}(\bk)$ and associated eigenfunctions are both $\Lambda^*$-periodic functions of $\bk$. %This can be checked by substituting $\bk$ with $\bk + \bmu$ where $\bmu \in \Lambda^*$, and \eqref{FB} remains unaltered.

     \item[iii)] We identify $\Psi_m(\bx,\bk)$ with its pseudo-periodic extension from $\bx \in \mathcal C$ to $\bx \in \RR^d$ by the boundary condition in $\bx$ in \eqref{FB}. Then for any function $\phi(\bx) \in L^2(\RR^d)$, we define its Bloch transform as
\begin{equation} \label{1220}
    \tilde{\phi}_m(\bk) = \int_{\RR^d} \phi(\bx) \overline{\Psi_{m}(\bx,\bk)} \,\rd \bx,
\end{equation}
then $ \tilde{\phi}_m(\bk)$ possesses the following properties \cite[pp. 484]{BFPR1999JSP}:
\begin{itemize}
    \item For $\bx \in \RR^d$, one has the following Bloch decomposition of $\phi(\bx)  \in L^2(\RR^d)$,
    \begin{align} \label{1221}
      \phi(\bx) = \frac{1}{|\mathcal{B}|} \sum_{m=1}^{\infty} \int_{\mathcal{B}} \tilde{\phi}_m(\bk) \Psi_{m} (\bx,\bk) \,\rd \bk.  
    \end{align}

    \item For $\phi(\bx), \psi(\bx) \in L^2(\RR^d)$,
    $$ \int_{\RR^d} \phi(\bx) \overline{\psi(\bx)} \,\rd \bx = \frac{1}{|\mathcal{B}|} \sum_{m=1}^{\infty} \int_{\mathcal{B}} \tilde{\phi}_m(\bk) \overline{ \tilde{\psi}_m(\bk)} \,\rd \bk.$$ 

    \item The mapping $\phi \mapsto \tilde{\phi}$ is one to one and onto from $L^{2}(\RR^d) \mapsto \oplus_{m} L^2(\mathcal{B})$.
\end{itemize}

    \item[iv)] Following from \eqref{1220} and \eqref{1221}, we can deduce additional orthogonality conditions: 
    \begin{equation*}
     \frac{1}{|\mathcal{B}|} \sum_{m=1}^{\infty}  \int_{\mathcal{B}} \Psi_{m}(\bx,\bk), \overline{\Psi_{m}(\by,\bk)} \,\rd \bk = \delta(\by - \bx),
    \end{equation*}
and
    \begin{equation} \label{25}
    \frac{1}{|\mathcal{B}|} \int_{\RR^d} \Psi_{j}(\bx, \bk), \overline{\Psi_{m} (\bx, \tilde{\bk})} \,\rd \bx  =  \delta_{jm}  \delta_{\text{per}}( \bk - \tilde{\bk} ),
    \end{equation}
where $\delta_{mj}$ is the Kronecker delta, and  $\delta_{\text{per}}$ is defined in the sense that, 
\[
\forall \, \varphi(\bk) \in C^{\infty}(\mathcal{B}), \quad \varphi(\bk) = \int_{\mathcal{B}} \varphi(\tilde{\bk}) \delta_{\text{per}}(\bk- \tilde{\bk}) \,\rd \tilde{\bk}.
\]
% for $\bk \in \mathcal{B}$ and $\tilde \bk \in \RR^d$, 
% \[
% \delta_{\text{per}}( \bk - \tilde{\bk} )  = \delta(\bk - (\tilde \bk + \bmu)), \quad ~ \text{where}~~ \bmu\in \Lambda^*, ~  \tilde \bk + \bmu \in \mathcal B\,.
% \]
\end{itemize}
 
%For each $\bk$, this problem is self-adjoint with compact resolvent, and hence there exist 
%Equation \eqref{FB} is commonly referred to as the \textit{Bloch} eigenvalue problem, and the associated eigenfunction $\Psi_m(\bx,\bk)$ is known as the Bloch state.  In a more general case, there may be a set of functions $\Psi_m^{\alpha}(\bx,\bk)$ with $\alpha = 1,...,r_m$ as the eigenfunctions corresponding to the eigenvalue $E_m(\bk)$. However, for the sake of notation simplicity, we set $r_m \equiv 1$ in this context. \lw{Please add references [2,13, 22] in Bal's paper here.} We refer the readers for more details about the Bloch theory in \cite{BLG1978, OK1964, Wilcox1978}.

%--------------------------------------------------------------------

\subsection{The Wigner Transform}
We now examine the Schr\"{o}dinger equation in the semi-classical scaling and introduce the Wigner transform, a crucial tool in our subsequent derivation. 

The semi-classical scaling of \eqref{s} with initial condition $\phi(t=0,\bx) = \phi^0(\bx)$ is expressed as:
\begin{equation}\label{ss}
    \left\{
    \begin{aligned}
    &\im \ve \frac{\partial \phi_{\ve}(t,\bx)}{\partial t} + \frac{\ve^2}{2} \Delta \phi_{\ve}(t,\bx) - V\left( \frac{\bx}{\ve} \right) \phi_{\ve}(t,\bx) = 0, \quad t \in \RR_{+}, \quad \bx \in \RR^d, \\
    & \phi_{\ve}(t=0,\bx) =\ \phi^0_{\ve}(\bx)\,.
    \end{aligned}
    \right.
\end{equation}
We define the \textit{(asymmetric) Wigner transform} of the solution $\phi_\ve(t,\bx)$ as:
\begin{equation}\label{Wigner}
    W_{\ve}(t,\bx,\bk) = \frac{1}{(2\pi)^d} \int_{\RR^d} \e^{-\im \bk \cdot \by} \phi_{\ve}(t,\bx-\ve \by) \, \overline{\phi_{\ve}(t,\bx)} \,\rd \by\,,
\end{equation}
then \eqref{ss} becomes:
\begin{multline} \label{WignerEq}
    \partial_t W_{\ve}(t,\bx, \bk) + \bk \cdot \nabla_\bx W_{\ve}(t,\bx, \bk) + \frac{\im \ve}{2} \Delta_\bx W_{\ve}(t,\bx, \bk)\\[4pt]
    = \frac{1}{\im \ve} \sum_{\bmu \in \Lambda^*} \e^{\im \bmu \cdot \frac{\bx}{\ve}} \hat{V}(\bmu) \left[ W_{\ve}(t,\bx,\bk-\bmu) - W_{\ve}(t,\bx,\bk) \right]\,,
\end{multline}
with the initial condition:
\begin{equation}\label{WignerEq-initial}
    W_{\ve}(t=0,\bx,\bk) = W_{\ve}^0(\bx,\bk) = \frac{1}{(2\pi)^d} \int_{\RR^d} \e^{-\im \bk \cdot \by} \phi^0_{\ve}(\bx-\ve \by) \, \overline{\phi^0_{\ve}(\bx)} \,\rd \by\,,
\end{equation}
where $\bmu \in \Lambda^*$ and 
\[
\hat V(\bmu) = \frac{1}{|\mathcal C|}\int_{\mathcal C} \e^{-\im \bmu \cdot \by} V(\by) \,\rd \by\,.
\]

% Then the energy and energy flux can be defined through $ W_{\ve}(t,\bx,\bk)$ via the following formulas: 
% \begin{align*}
%     E_{\ve}(t,\bx) &= |\phi_{\ve}(t,\bx)|^2 = \int_{\RR^d} W_{\ve}(t,\bx, \bk) \,\rd \bk \,, \\ 
%     F_{\ve}(t,\bx) &= \frac{1}{2\im} \left[\phi_{\ve}(t,\bx) \nabla_{\bx} \overline{\phi_{\ve}(t,\bx)} - \overline{\phi_{\ve}(t,\bx)} \nabla_{\bx} \phi_{\ve}(t,\bx)\right] = \int_{\RR^d} \bk W_{\ve}(t,\bx, \bk) \,\rd \bk\,.
% \end{align*}
% Additionally, the relation between $\nabla_{\bx}\phi_{\ve}(t,\bx) $ and the second moments in $\bk$ is given by:
% \begin{equation*}
%     \left| \nabla_{\bx}\phi_{\ve}(t,\bx) \right|^2 = \int_{\RR^d} |\bk|^2 W_{\ve}(t,\bx, \bk) \,\rd \bk\,.
% \end{equation*}
Hence, the Wigner transform provides a convenient approach for studying energy propagation in the phase space $ (\bx, \bk) \in \RR^d \times \RR^d$. See \cite{GMMP1997, LP1993, RPK1997, gerard1993ergodic} for additional insights into the properties of the Wigner transform.

\begin{remark}
    In addition to the asymmetric Wigner transform \eqref{Wigner}, one can alternatively use the symmetric version of the Wigner transform:
    \begin{equation*}
    \tilde{W}_{\ve}(t,\bx,\bk) = \frac{1}{(2\pi)^d} \int_{\RR^d} \e^{-\im \bk \cdot \by} \phi_{\ve}(t, \bx - \frac{\ve \by}{2}) \, \overline{\phi_{\ve}(t, \bx + \frac{\ve \by}{2})} \,\rd \by\,,
    \end{equation*}
    which is equivalent to $W_{\ve}(t,\bx,\bk)$ in the sense that they have the same weak limit as $\ve \downarrow 0$ \cite{BFPR1999JSP}. In the subsequent discussion, we will employ the asymmetric version as it yields a more concise result like \eqref{sigma_deterministic}. For further distinctions, one may refer to the discourse in \cite[pp.~515]{CJL2013KRM}.
\end{remark}

%%%%%%%%%%%%%%%%%%%%%%%%%%%%%%%%%%%%%%%%%%%%%%%%%%%%%%%%%%%

\section{Semi-classical model in a periodic structure without randomness}
\label{sec:without_randomness}

\subsection{Derivation of effective model by asymptotic analysis}
In this section, we explore the semi-classical limit of the Schr\"{o}dinger equation in the absence of randomness. Without loss of generality, we specifically focus on the case where two energy bands intersect. Our derivation closely aligns with the approach outlined in \cite{CJL2013KRM}, but extends it to arbitrary dimensions with any lattice structure. Our derivation also aligns with the method of \cite{BFPR1999JSP}, which addressed the case without band crossings.

We start by seeking a solution of \eqref{WignerEq} with the two-scale form:
\begin{equation*}
    W_\ve(t,\bx,\bk) \mapsto \left. W_\ve(t,\bx,\bz,\bk) \right|_{\bz = \frac{\bx}{\ve}},
\end{equation*}
where we abuse notation to write $W_\ve$ for both the original Wigner transform and its two-scale form. We will refer to $\bz = \frac{\bx}{\ve}$ as the fast variable. Replacing 
\[
\nabla_{\bx} \mapsto \nabla_{\bx} + \frac{1}{\ve} \nabla_{\bz}\,,
\]
in \eqref{WignerEq} to obtain:
% Denoting the solution to the semiclassical Schr\"odinger equation under the Wigner transform \eqref{WignerEq} as $W_{\ve}(t,\bx,\bz, \bk)$, we can rewrite \eqref{WignerEq} as follows:
\begin{multline}\label{WignerEqz}
    \partial_t W_{\ve}(t,\bx,\bz,\bk) + \bk \cdot \nabla_\bx W_{\ve}(t,\bx,\bz,\bk) + \frac{1}{\ve}\bk \cdot \nabla_\bz W_{\ve}(t,\bx,\bz,\bk) \\[4pt]
    + \frac{\im \ve}{2} \Delta_\bx W_{\ve}(t,\bx,\bz,\bk) +
     \im \nabla_\bz \cdot \nabla_\bx W_{\ve}(t,\bx,\bz,\bk) + \frac{\im}{2\ve} \Delta_{\bz} W_{\ve}(t,\bx,\bz,\bk) \\[4pt]
    = \frac{1}{\im \ve} \sum_{\bmu \in \Lambda^*} \e^{\im \bmu \cdot \bz} \hat{V}(\bmu) \left[ W_{\ve}(t,\bx,\bz,\bk-\bmu) - W_{\ve}(t,\bx,\bz,\bk) \right].
\end{multline}
We seek a solution by the formal asymptotic expansion: %of $W_{\ve}(t,\bx,\bz,\bk)$:
\begin{equation}\label{expansion1}
    W_{\ve}(t,\bx,\bz,\bk) = W_0(t,\bx,\bz,\bk) + \ve W_1 (t,\bx,\bz,\bk) + \ve^2 W_2 (t,\bx,\bz,\bk) + ....
\end{equation}
Substituting into \eqref{WignerEqz} and equating terms of like order in $\ve$, we obtain a sequence of equations.

\textbf{(I) At order $O\left( \frac{1}{\ve} \right)$}, we have:
\begin{equation}\label{Order-1}
    \underbrace{ \bk \cdot \nabla_\bz W_0 + \frac{\im}{2} \Delta_\bz W_0 - \frac{1}{\im} \sum_{\bmu \in \Lambda^*} \e^{\im \bmu \cdot \bz} \hat{V}(\bmu) [W_0(t,\bx,\bz,\bk-\bmu) - W_0(t,\bx,\bz,\bk)]}_{ \eqdef  \mathcal{L}[W_0](t,\bx,\bz,\bk),} = 0\,,
\end{equation}
where we have defined the operator $\mathcal{L}$, which is skew-symmetric with respect to the inner product \eqref{inner_prod}. For \eqref{Order-1} to be satisfied, $W_0(t,\bx,\bz,\bk)$ must belong to $\text{Ker} \mathcal{L}$. 
%where $\mathcal{L}$ is a skew-symmetric operator.  This indicates that $W_0(t, \bx, \bz, \bk)$ belongs to  

To characterize $\text{Ker} \mathcal{L}$, note first that any $\bk \in \RR^d$ can be uniquely decomposed into 
\begin{equation} \label{kpmu}
    \bk = \bp + \bmu, \quad \text{with} \quad \bp \in \mathcal{B} \quad \text{and} \quad \bmu \in \Lambda^*\,.
\end{equation}
For positive integers $m, n$, we can define  $Q_{mn}(\bz, \bmu, \bp)$ using  the Bloch functions as:
\begin{equation}\label{Qmn}
    Q_{mn}(\bz, \bmu, \bp) = \frac{1}{|\mathcal{C}|} \int_{\mathcal{C}} \e^{\im (\bp + \bmu)\cdot \by} \Psi_m (\bz- \by, \bp) \overline{\Psi_n(\bz, \bp)} \,\rd \by\,.
\end{equation}
Then $ Q_{mn}(\bz, \bmu, \bp) $ is $\Lambda$-periodic in $\bz$ and satisfies: 
\begin{equation*}
    \mathcal{L} [Q_{mn}](\bz, \bmu, \bp) = \im \big(E_m(\bp) - E_n(\bp)\big) Q_{mn}(\bz, \bmu, \bp).
\end{equation*}
Clearly, we have $Q_{mm} \in \text{Ker} \mathcal{L}$ for all $m$. For $\bp_*$ such that $E_m(\bp_*) = E_n(\bp_*)$ for $m \neq n$, i.e., at band crossings (for example, Dirac points in graphene; see Section \ref{sec:application}), $Q_{mn}$ also belongs to $\text{Ker} \mathcal{L}$. %Such a $\bp_*$ is termed the Dirac point. 

In order to describe effects arising from band crossings, following \cite{CJL2013KRM}, we take $W_0(t, \bx, \bz, \bk)$ as:
\begin{equation} \label{W0}
    W_0(t, \bx, \bz, \bk) = \sum_{m,n = 1}^2 \sigma_{mn}(t, \bx, \bp)  Q_{mn}(\bz, \bmu, \bp),
\end{equation}
where $\bk \in \RR^d$, $\bp \in \mathcal B$ and $\bmu \in \Lambda^*$ are related as described in \eqref{kpmu} and $\sigma_{mn}$ is called the coherence function. 

\begin{remark}
    A more general expansion of $W_0$ is:
\begin{equation}
    W_0(t,\bx,\bz,\bk) = \sum_{m,n,\alpha,\beta} \sigma_{mn}^{\alpha \beta} (t,\bx,\bp) Q_{mn}^{\alpha \beta}(\bz,\bmu,\bp)\,,
\end{equation}
where the superscripts $\alpha, \beta$ label the eigenfunctions inside the eigenspace. Since our main focus is on handling band crossings, we assume the multiplicity of each eigenvalue $E_m(\bp)$ is $1$ for simplicity (this assumption is known to be true for the leading eigenvalue when the Fourier transform of the periodic potential $V$ is negative \cite{BFPR1999JSP}), so that no superscripts is needed in the asymptotic expansion form \eqref{W0}. Higher multiplicity will not bring essential difficulties in our following calculation.
\end{remark}

We assume that $\phi_\ve^0$ in \eqref{ss} is such that the Wigner-transformed initial condition in \eqref{WignerEq-initial} can be expressed as
\begin{equation} \label{eq:init_W}
    W_\ve(0,\bx,\bk) = \left. \sum_{m,n = 1}^2 \sigma^0_{mn}(\bx,\bp) Q_{mn}(\bz,\bmu,\bp) \right|_{\bz = \frac{\bx}{\ve}}
\end{equation}
for some $\sigma_{mn}^0$. It is important to note that $W_0$ in the form of \eqref{W0} does not always belong to $\text{Ker} \mathcal{L}$, so that \eqref{Order-1} is not exactly satisfied. Instead, we have
\begin{equation} \label{eq:this}
\begin{split}
    \mathcal L [W_0] =& \sum_{m,n=1}^2 \sigma_{mn}(t,\bx,\bp) \mathcal L [Q_{mn}(\bz,\bmu,\bp)] \\
    =& \, \sigma_{12}(t,\bx,\bp) \mathcal L [Q_{12}(\bz,\bmu,\bp)] + \sigma_{21}(t,\bx,\bp) \mathcal L [Q_{21}(\bz,\bmu,\bp)],
\end{split}
\end{equation}
and for $\bp \neq \bp_*$, i.e., away from band crossings, the right-hand side is non-zero. Nevertheless, we still use the form \eqref{W0} as it encodes the interaction between bands as we will see next.

%\lw{I don't quite understand the point of this remark. } \kq{just to exclude the case $E_m(\bp) = E_{n}(\bp)$ for all $\bp$. Because in this case, it is nonsense to discuss the evolution of $\sigma_{12} $ and $\sigma_{21}$. This is not an essential remark I think.} \lw{is (3.9) the correct equation to cite here?}
%\kq{Yes, (3.9) should be correct. Due to 
% \[
% \mathcal{L} [Q_{mn}](\bz, \bmu, \bp) = \im \big(E_m(\bp) - E_n(\bp)\big) Q_{mn}(\bz, \bmu, \bp).
% \]
% the right-hand side of will go vanishment, if $E_1(\bp) = E_2(\bp)$ for all $\bp$. Then, it would be nonsense to talk about $\sigma_{12}$ and $\sigma_{21}$}
\begin{remark}
    Note that the right-hand side of \eqref{eq:this} will be zero for all $\bp$ if the bands $E_m$ and $E_n$ coincide everywhere, i.e., are everywhere degenerate. We ignore this case in the present work since we do not anticipate interesting new phenomena arising from this generalization. Our results will generalize, but be more complicated. For example, the coherence functions $\sigma_{mn}$ appearing in \eqref{W0} must become matrix-valued.
\end{remark}
%Here, $\sigma_{mn}$ is the so-called coherence matrices that are defined inside the Brillouin zone $\bp \in \mathcal{B}$ and can be generalized to the whole space $\RR^d$ by the $\Lambda^*$-periodicity in $\bp$.

%-------------------------------------------

\textbf{(II) At order $O(1)$}, we have 
\begin{equation}\label{order1}
    \frac{\partial W_0}{\partial t} + \bk \cdot \nabla_\bx W_0 + \im \nabla_\bx \cdot \nabla_\bz W_0 = - \mathcal{L}[W_1] - \frac{1}{\ve} \mathcal{L}[W_0].
\end{equation}
Again, $\frac{1}{\ve} \mathcal{L}[W_0]$ needs to be kept because $\mathcal{L}[W_0]$ is not always zero when $W_0$ takes the form of \eqref{W0}.  

To proceed with the derivation, we first introduce the following orthogonal relations between $Q_{mn}$.
Let $\left< \cdot \,,  \cdot \right>_{\mathcal{C},\Lambda^*}$ denote integration in $\bz$ over the fundamental cell $\mathcal{C}$ and summation over $\bmu \in \Lambda^*$, i.e.,
\begin{equation} \label{inner_prod}
    \left< f(\cdot, \cdot, \bp) \,,  g(\cdot, \cdot, \bp) \right>_{\mathcal{C},\Lambda^*} := \ \sum_{\bmu \in \Lambda^*} \frac{1}{|\mathcal{C}|} \int_{\mathcal{C}} f(\bz, \bmu, \bp) \,\overline{g(\bz, \bmu, \bp)}  \,\rd \bz\,,
\end{equation}
and $\left< \cdot \,,  \cdot \right>_{\mathcal{C}}$ be the integration in  $\bz$ over the fundamental cell $\mathcal{C}$
\begin{equation*}
    \left< f(\cdot,  \bp) \,,  g(\cdot,  \bp) \right>_{\mathcal{C}} := \ \frac{1}{|\mathcal{C}|} \int_{\mathcal{C}} f(\bz, \bp) \,\overline{g(\bz, \bp)}  \,\rd \bz.
\end{equation*}
Then, we have, for $\bp \in \mathcal{B}$,
    \begin{equation}\label{Qmnjl}
    \begin{split}
        \left<  Q_{mn}(\cdot, \cdot, \bp) \,, Q_{jl}(\cdot, \cdot, \bp) \right>_{\mathcal{C},\Lambda^*} =& \ \sum_{\bmu \in \Lambda^*} \frac{1}{|\mathcal{C}|} \int_{\mathcal{C}} Q_{mn}(\bz, \bmu, \bp) \,\overline{Q_{jl}(\bz, \bmu, \bp)}  \,\rd \bz \\[4pt]
        =& \  \delta_{m j} \delta_{n l}\,;
    \end{split}
    \end{equation}
and for $\bk = \bp + \bmu$,
    \begin{equation}\label{kQmnjl}
    \begin{split}
        \left<  \bk Q_{mn}(\cdot, \cdot, \bp)  \,, Q_{jl}(\cdot, \cdot, \bp)  \right>_{\mathcal{C},\Lambda^*} = & \ \sum_{\bmu \in \Lambda^*} \frac{1}{|\mathcal{C}|} \int_{\mathcal{C}} (\bp + \bmu) Q_{mn}(\bz, \bmu, \bp) \,\overline{Q_{jl} (\bz, \bmu, \bp)} \,\rd \bz \\[4pt]
        =& \  \left< (- \im \nabla_z) \Psi_m(\cdot, \bp)  \,, \Psi_{j}(\cdot, \bp)  \right>_{\mathcal{C}} \delta_{n l}\,;
    \end{split}
    \end{equation}
and 
\begin{equation}\label{gQmnjl}
\begin{split}
    &  \left<  \nabla_\bz Q_{mn}(\cdot, \cdot, \bp) \,,\, Q_{jl}(\cdot, \cdot, \bp) \right>_{\mathcal{C},\Lambda^*} 
    \\ = &
    \sum_{\bmu \in \Lambda^*} \frac{1}{|\mathcal{C}|} \int_{\mathcal{C}}  \nabla_\bz Q_{mn}(\bz, \bmu, \bp) \,\overline{Q_{jl} (\bz, \bmu, \bp)} \,\rd \bz \\[4pt]
    = &- \left<  \nabla_\bz \Psi_m(\cdot, \bp) \,, \Psi_{j}(\cdot, \bp) \right>_{\mathcal{C}} \,\delta_{n l} + \overline{ \left<  \nabla_\bz \Psi_n(\cdot, \bp) \,, \Psi_{l}(\cdot, \bp) \right>_{\mathcal{C}} } \,\delta_{m j}\,.
\end{split}
\end{equation}
We refer to \cite[pp.~486]{BFPR1999JSP} for further details, and additional calculations are provided in Appendix \ref{appendix-A} for completeness.

Returning to \eqref{order1}, we can obtain the equation for $W_0$ by imposing a solvability condition on the equation for $W_1$. Specifically, taking inner product on both sides of \eqref{order1} with  $Q_{jl} ~ ( j, l \in \{1,2\} )$ in $\bz$ and $\bmu$, we have 
\begin{equation*}
\begin{split}
    &\sum_{m, n=1}^2 \left[  \left< \partial_t \sigma_{mn} Q_{mn} \,, Q_{jl} \right>_{\mathcal{C},\Lambda^*}  +   \left<  \bk  \cdot \nabla_\bx \sigma_{mn} Q_{mn} \,, Q_{jl}  \right>_{\mathcal{C},\Lambda^*}  +   \left<  \im \nabla_\bz \cdot\nabla_\bx \sigma_{mn} Q_{mn} \,, Q_{jl}  \right>_{\mathcal{C},\Lambda^*} \right] \\
    =& \left< \mathcal{L}[W_1] \,, Q_{jl} \right>_{\mathcal{C},\Lambda^*} - \frac{1}{\ve}  \sum_{m,n=1}^2  \left<  \sigma_{mn} \mathcal{L}[Q_{mn}] \,, Q_{jl}  \right>_{\mathcal{C},\Lambda^*} \, ,
\end{split}
\end{equation*}
which is further simplified as follows:
\begin{equation}\label{order1-sigma-full}
\begin{split}
    &\sum_{m, n=1}^2 \left[ \partial_t \sigma_{mn} \left< Q_{mn} \,, Q_{jl} \right>_{\mathcal{C},\Lambda^*}  +  \nabla_\bx \sigma_{mn} \cdot \left<  \bk Q_{mn} \,, Q_{jl}  \right>_{\mathcal{C},\Lambda^*}  +  \im \nabla_\bx \sigma_{mn} \cdot \left< \nabla_\bz Q_{mn} \,, Q_{jl}  \right>_{\mathcal{C},\Lambda^*} \right] \\
    =& -\left< W_1 \,, \mathcal{L}[Q_{jl}] \right>_{\mathcal{C},\Lambda^*} - \sum_{m,n=1}^2 \sigma_{mn} \frac{\im}{\ve} [E_m(\bp) - E_n(\bp)]  \left< Q_{mn} \,, Q_{jl}  \right>_{\mathcal{C},\Lambda^*} \\
    =& \im [E_l(\bp) - E_j(\bp)] \left< W_1 \,, Q_{jl} \right>_{\mathcal{C},\Lambda^*} - \sum_{m,n=1}^2 \sigma_{mn} \frac{\im}{\ve} [E_m(\bp) - E_n(\bp)]  \left< Q_{mn} \,, Q_{jl} \right>_{\mathcal{C},\Lambda^*}\,.
\end{split}
\end{equation}
Note that the term $\im [E_l(\bp) - E_j(\bp)] \left< W_1 \,, Q_{jl} \right>_{\mathcal{C},\Lambda^*}$ can be considered negligible. The rationale is as follows: when $\bp = \bp_*$, where $\bp_*$ is the band crossings point, i.e., $E_1(\bp_*) = E_2(\bp_*)$, this term becomes zero. In a small neighborhood of $\bp_*$, where $\bp$ is close to $\bp_*$, the quantity $|E_1(\bp) - E_2(\bp)|$ is much smaller than 1. Furthermore, when $\bp$ is far away from $\bp_*$ and the two bands are well separated, one should follow the derivation in \cite{BFPR1999JSP} for well-separated energy bands, where this term no longer appears. For these reasons, we drop this term in what follows.

By substituting the orthogonality conditions \eqref{Qmnjl}, \eqref{kQmnjl} and \eqref{gQmnjl}, we can simplify \eqref{order1-sigma-full} into: 
% \begin{equation}
%     \partial_t \sigma_{jl} + \sum_{n=1}^2 \nabla_\bx \sigma_{j n} \cdot \overline{  \left<  (-\im \nabla_\bz) \Psi_{n} \,, \Psi_l \right>_{\bz} }  =  \frac{\im}{\ve} [E_l(\bp) - E_j(\bp)] \sigma_{jl} 
% \end{equation}
% or
\begin{equation} \label{eq:W_Dirac}
\partial_t \sigma_{jl} + \sum_{n=1}^{2}  \nabla_\bx \sigma_{j n} \cdot \left<  (-\im \nabla_\bz) 
 \Psi_{l} \,, \Psi_n \right>_{\mathcal{C}} = \frac{\im}{\ve} [E_l(\bp) - E_j(\bp)] \sigma_{jl}, \quad j,l \in \{1,2\},
\end{equation}
which must be solved subject to the initial condition
\begin{equation*}
    \sigma_{jl}(0,\bx,\bp) = \sigma^0_{jl}(\bx,\bp), \quad j,l \in \{1,2\},
\end{equation*}
where $\sigma^0_{jl}$ are as in \eqref{eq:init_W}. We refer to the term on the right-hand side of \eqref{eq:W_Dirac} as a relaxation-type term because of its similarity to terms appearing in the models of kinetic theory and conservation law; see \cite{BGK1954, JinXin1995}. The system \eqref{eq:W_Dirac} can be written in the following matrix form: 
\begin{multline}\label{sigma_deterministic}
    \partial_t 
    \begin{pmatrix} 
    \sigma_{11} & \sigma_{12} \\ 
    \sigma_{21} & \sigma_{22} 
    \end{pmatrix} 
    + \nabla_\bx 
    \begin{pmatrix} 
    \sigma_{11} & \sigma_{12} \\ 
    \sigma_{21} & \sigma_{22} 
    \end{pmatrix}
    \cdot  
    \begin{pmatrix} 
     \left< (-\im \nabla_\bz) \Psi_1 \,,  \Psi_1 \right>_{\mathcal{C}}  &  \left< (-\im \nabla_\bz) \Psi_2 \,,  \Psi_1 \right>_{\mathcal{C}}\\[4pt]
    \left< (-\im \nabla_\bz) \Psi_1 \,, \Psi_2 \right>_{\mathcal{C}} & \left< (-\im \nabla_\bz) \Psi_2 \,,  \Psi_2 \right>_{\mathcal{C}}
    \end{pmatrix}
     \\[8pt]
    = \frac{\im}{\ve}
    \begin{pmatrix} 
    0 & [E_2(\bp)-E_1(\bp)] \sigma_{12} \\ 
    [E_1(\bp)-E_2(\bp)] \sigma_{21} & 0
    \end{pmatrix},
\end{multline}
where we have introduced the convenient shorthand notation
\begin{equation*}
    \begin{split}
    &\nabla_\bx 
    \begin{pmatrix} 
    \sigma_{11} & \sigma_{12} \\ 
    \sigma_{21} & \sigma_{22} 
    \end{pmatrix}
    \cdot  
    \begin{pmatrix} 
     \left< (-\im \nabla_\bz) \Psi_1 \,,  \Psi_1 \right>_{\mathcal{C}}  &  \left< (-\im \nabla_\bz) \Psi_2 \,,  \Psi_1 \right>_{\mathcal{C}}\\
    \left< (-\im \nabla_\bz) \Psi_1 \,, \Psi_2 \right>_{\mathcal{C}} & \left< (-\im \nabla_\bz) \Psi_2 \,,  \Psi_2 \right>_{\mathcal{C}}
    \end{pmatrix}
    :=  \\
    &\quad \quad \quad \quad \sum_{i = 1}^d 
    \begin{pmatrix} 
        \partial_{x_i} \sigma_{11} & \partial_{x_i} \sigma_{12} \\ 
    \partial_{x_i} \sigma_{21} & \partial_{x_i} \sigma_{22} 
    \end{pmatrix}
       \begin{pmatrix} 
           \left< (-\im \partial_{z_i}) \Psi_1 \,,  \Psi_1 \right>_{\mathcal{C}} & \left< (-\im \partial_{z_i}) \Psi_2 \,,  \Psi_1 \right>_{\mathcal{C}} \\
           \left< (-\im \partial_{z_i}) \Psi_1 \,, \Psi_2 \right>_{\mathcal{C}} & \left< (-\im \partial_{z_i}) \Psi_2 \,,  \Psi_2 \right>_{\mathcal{C}}
    \end{pmatrix}.
    \end{split}
\end{equation*}
We can write \eqref{sigma_deterministic} more concisely by introducing
 \begin{equation*}
 \sigma := 
     \begin{pmatrix} 
     \sigma_{11} & \sigma_{12} \\ \sigma_{21} & \sigma_{22} 
     \end{pmatrix},
     \quad  D := 
     \begin{pmatrix} 
      \left< (-\im \nabla_\bz) \Psi_1 \,,  \Psi_1 \right>_{\mathcal{C}}  &  \left< (-\im \nabla_\bz) \Psi_2 \,,  \Psi_1 \right>_{\mathcal{C}}\\[4pt]
     \left< (-\im \nabla_\bz) \Psi_1 \,, \Psi_2 \right>_{\mathcal{C}} & \left< (-\im \nabla_\bz) \Psi_2 \,,  \Psi_2 \right>_{\mathcal{C}}
     \end{pmatrix}, 
 \end{equation*}
 \begin{equation*}
     E : = \frac{\im}{\ve}
     \begin{pmatrix} 
     0 & [E_2(\bp)-E_1(\bp)] \sigma_{12} \\ 
     [E_1(\bp)-E_2(\bp)] \sigma_{21} & 0 
     \end{pmatrix}, \quad \sigma^0 := 
     \begin{pmatrix} 
     \sigma^0_{11} & \sigma^0_{12} \\ \sigma^0_{21} & \sigma^0_{22} 
     \end{pmatrix}.
 \end{equation*}
     The system \eqref{sigma_deterministic} can then be presented as
 \begin{equation*}
     \partial_t \sigma + \nabla_\bx \sigma \cdot D = E, \quad \sigma(0) = \sigma^0.
 \end{equation*}

 \begin{remark}
     Noting that $\overline{ D^T } = D$, taking the complex conjugate and transpose of \eqref{sigma_deterministic} we see that $\overline{\sigma^T}$ satisfies an equivalent system to \eqref{sigma_deterministic} where $D$ acts by matrix multiplication from the left
 \begin{equation*}
     \partial_t \overline{ \sigma^T } + \overline{D^T} \cdot \nabla_x \overline{ \sigma^T } = \partial_t \overline{ \sigma^T } + D \cdot \nabla_x \overline{ \sigma^T } = \overline{E^T}.
 \end{equation*}
     %Note that by redefining $\sigma$, the system \eqref{sigma_deterministic} is equivalent to a system where the 
     %In fact, there is an alternative way to represent the system \eqref{sigma_deterministic}. Let
 %\begin{equation}
     %D := 
     %\begin{pmatrix} 
      %\left< (-\im \nabla_\bz) \Psi_1 \,,  \Psi_1 \right>_{\bz}  &  \left< (-\im \nabla_\bz) \Psi_2 \,,  \Psi_1 \right>_{\bz}\\[4pt]
     %\left< (-\im \nabla_\bz) \Psi_1 \,, \Psi_2 \right>_{\bz} & \left< (-\im \nabla_\bz) \Psi_2 \,,  \Psi_2 \right>_{\bz}
     %\end{pmatrix},
 %\end{equation}
 %and note that $\overline{ D^T } = D$. Let
 %\begin{equation}
     %\sigma := 
     %\begin{pmatrix} 
     %\sigma_{11} & \sigma_{12} \\ \sigma_{21} & \sigma_{22} 
     %\end{pmatrix},
 %\end{equation}
 %and
 %\begin{equation}
     %E : = \frac{\im}{\ve}
     %\begin{pmatrix} 
     %0 & [E_2(\bp)-E_1(\bp)] \sigma_{12} \\ 
     %[E_1(\bp)-E_2(\bp)] \sigma_{21} & 0 
     %\end{pmatrix}.
 %\end{equation}
 %Then, we have that
 %\begin{equation}
     %\partial_t \sigma + \nabla_\bx \sigma \cdot D = E.
 %\end{equation}
 %Taking the transpose and complex conjugate, we have
 %\begin{equation}
     %\partial_t \overline{ \sigma^T } + \overline{D^T} \cdot \nabla_x \overline{ \sigma^T } = \partial_t \overline{ \sigma^T } + D \cdot \nabla_x \overline{ \sigma^T } = \overline{E^T}
 %\end{equation}
 %%\partial_t \overline{ \sigma^T } + \overline{D^T} \cdot \nabla_\bx \overline{ \sigma^T } = \partial_t \overline{ \sigma^T } + D \cdot \nabla_\bx \overline{ \sigma^T } = \overline{E^T},
 %% so that each column of
 %% \begin{equation}
 %%     \Sigma := \overline{ \sigma^T }
 %% \end{equation}
 %% satisfies the conventional case where the matrix acts from the left. 
 \end{remark}

%\subsection{\blue{Reduction of derived model in special cases}} 
\subsection{Alternative formulation and domain decomposition of the derived model}
\label{sec:reduction}

The preceding calculation shows that making the ansatz \eqref{W0} produces terms proportional to $\frac{1}{\ve}$ in equation \eqref{sigma_deterministic}. This idea for dealing with systems with crossings goes back to \cite{CJL2013KRM} but still demands further explanation and justification since these terms seem to invalidate the asymptotic expansion. We now show how the system \eqref{sigma_deterministic} reduces to the Liouville equation for $\bp$ when the bands are well separated, while still capturing the diabatic (band crossing) effect for $\bp$ when the bands are close or touching.

\begin{figure}[h]
    \centering
    \includegraphics[width=0.75\textwidth]{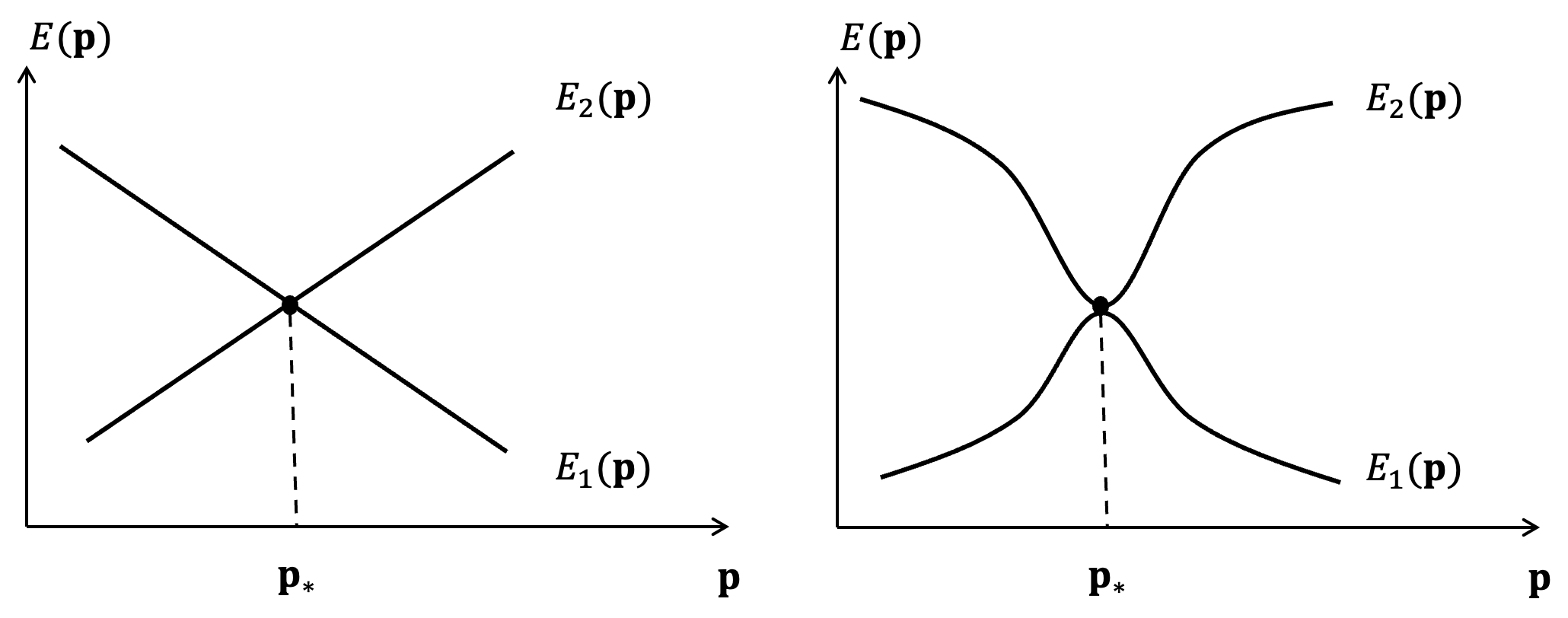}%{Submission/single-band-crossings.png}
    \caption{Left: A conical $(r = 1)$ energy band crossing. Right: A quadratic $(r = 2)$ energy band crossing.}
    \label{fig1}
\end{figure}

We start by introducing 
\begin{equation} \label{eq:transformation}
    \widetilde{\sigma}_{jl}(t,\bx,\bp) := \e^{\frac{\im}{\ve} [E_j(\bp) - E_l(\bp)] t} \sigma_{jl}(t,\bx,\bp), \quad j,l \in \{1,2\},
\end{equation}
and then re-write the system \eqref{sigma_deterministic} as a system for $\widetilde{\sigma}_{jl}$ as
\begin{equation}\label{sigma_tilde_deterministic}
    \partial_t \widetilde{\sigma}_{jl}(t,\bx,\bp) + \sum_{n=1}^{2}  \nabla_\bx \widetilde{\sigma}_{j n}(t,\bx,\bp) \cdot \e^{- \frac{\im}{\ve}[E_l(\bp) - E_n(\bp)]t} \left<  (-\im \nabla_\bz)  \Psi_{l} \,, \Psi_n \right>_{\mathcal{C}} = 0,
\end{equation}
or in matrix form, (clearly, $\widetilde{\sigma}_{jl} = \sigma_{jl}$ for $j = l$),
\begin{small}
    \begin{equation} 
    \begin{split}
        & \mathbf{0} = \partial_t 
    \begin{pmatrix} 
        \sigma_{11} & \widetilde{\sigma}_{12} \\ 
        \widetilde{\sigma}_{21} & \sigma_{22} 
    \end{pmatrix}   \\
        &+ \nabla_\bx 
    \begin{pmatrix} 
        \sigma_{11} & \widetilde{\sigma}_{12} \\ 
        \widetilde{\sigma}_{21} & \sigma_{22} 
    \end{pmatrix}
    \cdot  
    \begin{pmatrix} 
        \left< (-\im \nabla_\bz) \Psi_1 \,,  \Psi_1 \right>_{\mathcal{C}}  & \e^{- \frac{\im}{\ve}[E_2(\bp) - E_1(\bp)]t}  \left< (-\im \nabla_\bz) \Psi_2 \,,  \Psi_1 \right>_{\mathcal{C}}\\[4pt]
        \e^{- \frac{\im}{\ve}[E_1(\bp) - E_2(\bp)]t} \left< (-\im \nabla_\bz) \Psi_1 \,, \Psi_2 \right>_{\mathcal{C}} & \left< (-\im \nabla_\bz) \Psi_2 \,,  \Psi_2 \right>_{\mathcal{C}}
    \end{pmatrix}.
    \end{split}
\end{equation}
\end{small}
We can now extract the limiting behavior of solutions of \eqref{sigma_tilde_deterministic}, and hence of \eqref{sigma_deterministic}, in the limit $\ve \downarrow 0$ for different values of $\bp$. For simplicity, we assume at this point that the bands $E_1, E_2$ have finitely many separated crossing points $\bp_*$. The case where eigenvalues cross along a curve can be dealt with similarly. We then assume that in a neighborhood of each $\bp_*$ the crossing has the following structure 
\begin{equation} \label{eq:lin_cross}
    E_1(\bp) = - \lambda_\sharp |\bp - \bp_*|^r + o(|\bp - \bp_*|^r), \quad E_2(\bp) = \lambda_\sharp |\bp - \bp_*|^r + o(|\bp - \bp_*|^r), \quad \bp \rightarrow \bp_*,
\end{equation}
for some positive constants $\lambda_\sharp$ and positive integers $r > 0$. This assumption allows for both conical ($r = 1$) and quadratic ($r = 2$) crossings, see fig \ref{fig1}. The case $r = 1$ is the situation encountered in graphene, see \eqref{eq:conical}, while $r = 2$ occurs in untwisted and twisted bilayer graphene \cite{NGPNG2009, becker2024diracconesmagicangles}.

%\kq{I think this part about the interpretation for graphene can be moved to Section 5 in response to Reviewer 3. And I'll add a figure of the band-crossings $K$ and $K'$ for graphene.} \aw{We could make the discussion here more general and allow for crossings like $|p - p^*|^s$ here. Then the regime is $\ve^{1/s}$. But we need something like this here otherwise people like Schotland will think the $1/\ve$ terms are mistakes. It looks like our model diverges as $\ve \downarrow 0$ which is absurd. I am happy to add more "physics interpretation" for that reviewer in the later section.}
We then have the following dichotomy about the domain decomposition, depending on the nature of the crossing.
\begin{itemize}
    \item When $|\bp - \bp_*| \gg \ve^{1/r}$ as $\ve \downarrow 0$, the functions $\e^{- \frac{\im}{\ve}[E_l(\bp) - E_n(\bp)]}$ are highly oscillatory with mean zero as $\ve \downarrow 0$ for $l \neq n$. By standard arguments (see, e.g., \cite{2008allaire}), the solution of \eqref{sigma_tilde_deterministic} converges weakly to the solution of the following system 
    % the terms in \eqref{sigma_tilde_deterministic} coupling the diagonal entries $\sigma_{11}$ and $\sigma_{22}$ to the off-diagonal entries $\widetilde{\sigma}_{12}$ and $\widetilde{\sigma}_{21}$ are highly oscillatory with mean zero as $\ve \downarrow 0$. By standard arguments (see, e.g., \cite{2008allaire}), the solution of \eqref{sigma_tilde_deterministic} converges weakly to the solution of the system where these terms are replaced by $0$, 
    % In this case, the diagonal and off-diagonal $\sigma$'s evolve according to the decoupled equations
    \begin{equation}
        \partial_t \widetilde{\sigma}_{jl}(t,\bx,\bp) + \nabla_\bx \widetilde{\sigma}_{jl}(t,\bx,\bp) \cdot  \left< (-\im \nabla_\bz) \Psi_{l} \,, \Psi_l \right>_{\mathcal{C}} = 0, \quad j,l \in \{1,2\}\,,
    \end{equation}
    which is obtained by replacing the relaxation term $\e^{- \frac{\im}{\ve}[E_l(\bp) - E_n(\bp)]}$ in \eqref{sigma_tilde_deterministic} to $\delta_{ln}$.       
    When these equations are expressed in terms of the eigenvalue band functions we recover the familiar form of the Liouville equation
        \begin{equation}
            \partial_t \widetilde{\sigma}_{jl} + \nabla_{\bp} E_l(\bp) \cdot \nabla_{\bx} \widetilde{\sigma}_{jl} = 0, \quad j,l \in \{1,2\}.
        \end{equation}
    In particular, we recover the result of \cite{BFPR1999JSP}, since when $\widetilde{\sigma}_{12}$ and $\widetilde{\sigma}_{21}$ are initially zero, they remain zero for each positive $t$ in the limit $\ve \downarrow 0$, while $\sigma_{11}$ and $\sigma_{22}$ evolve according to the same Liouville equations as derived in \cite[Eq.~(2.24)]{BFPR1999JSP}.
    \item When $|\bp - \bp_*| = O(\ve^{1/r})$ as $\ve \downarrow 0$, the functions $\e^{- \frac{\im}{\ve}[E_l(\bp) - E_n(\bp)]}$ are no longer highly oscillatory and cannot be replaced by $\delta_{ln}$. In particular, in this regime the coupling between the diagonal entries $\sigma_{11}$ and $\sigma_{22}$ to the off-diagonal entries $\widetilde{\sigma}_{12}$ and $\widetilde{\sigma}_{21}$ is no longer $o(1)$ as $\ve \downarrow 0$, 
    reflecting the fact that in this regime interband coupling effects cannot be ignored. Note also that in this regime the terms $\frac{\im}{\ve}[ E_l(\bp) - E_j(\bp) ]$ in \eqref{sigma_deterministic} no longer blow up as $\ve \downarrow 0$.
\end{itemize}
We thus have that, even though the system \eqref{sigma_deterministic} involves terms proportional to $\frac{1}{\ve}$, its solutions make sense in every case and indeed reduce to previously known results as expected. We remark further that the terms proportional to $\frac{1}{\ve}$ in the expansion could be removed from the beginning of the calculation by setting, instead of \eqref{W0},
\begin{equation} \label{W0_2}
    W_0(t, \bx, \bz, \bk) = \sum_{m,n = 1}^2 \e^{- \frac{\im}{\ve} [E_m(\bp) - E_n(\bp)] t} \sigma_{mn}(t, \bx, \bp)  Q_{mn}(\bz, \bmu, \bp).
\end{equation}
Equating terms of order $O(1)$ in the expansion would then give \eqref{sigma_tilde_deterministic} directly. Since the models obtained are ultimately equivalent, and the calculations become somewhat more complicated in the random case if we make the ansatz \eqref{W0_2}, we prefer to make the ansatz \eqref{W0}.

%%%%%%%%%%%%%%%%%%%%%%%%%%%%%%%%%%%%%%%%%%%%%%%%%%%%%%%%%
\section{Semi-classical model in a periodic structure with randomness}
\label{sec:with_randomness}

We now consider a weak random perturbation of the problem \eqref{ss} as follows:
\begin{equation}\label{random_system}
    \left\{
    \begin{aligned}
    &\im \ve \frac{\partial \phi_{\ve}}{\partial t} + \frac{\ve^2}{2} \Delta_\bx \phi_{\ve} - V\left( \frac{\bx}{\ve} \right) \phi_{\ve} - \sqrt{\ve} N\left( \frac{\bx}{\ve} \right)  \phi_{\ve} =\  0, \\
    &\phi_{\ve}(t=0,\bx) =\ \phi^0_{\ve}(\bx)\,.
    \end{aligned}
    \right.
\end{equation}
Here $N$ is a stationary, real-valued mean zero, spatially homogeneous random function on the probability space $(\Omega, \mathcal F, \mathbb P)$ with covariance given by: 
\begin{equation}\label{NNR}
   \expec{[N(\by) N(\by+\bx)] } = R(\bx), \quad \expec{ [\hat{N}(\bp) \hat{N}(\bq)] }   = (2\pi)^d \hat{R}(\bq) \, \delta(\bp+\bq)\,,
\end{equation}
where $\hat{N}(\bq)$ is the Fourier transform of $N(\by)$:
\[
\hat N(\bq) = \int_{\RR^d} \e^{-\im \bq \cdot \by} N(\by) \,\rd \by\,.
\]
By the scaling of $N$ in \eqref{random_system}, we are considering a random inhomogeneity with correlation length comparable to the wavelength $\varepsilon$ and a small variance.

Upon applying the Wigner transform to both sides of \eqref{random_system}, we obtain 
\begin{multline}\label{Wve}
    \partial_t W_{\ve} + \bk \cdot \nabla_\bx W_{\ve} + \frac{\im \ve}{2} \Delta_\bx W_{\ve} = \frac{1}{\im \ve} \sum_{\bmu \in \Lambda^*} \e^{\im \bmu \cdot \bz} \hat{V}(\bmu) \left[ W_{\ve}(t,\bx,\bk-\bmu) - W_{\ve}(t,\bx,\bk) \right]\\[4pt]
     + \frac{1}{\im \sqrt{\ve}} \frac{1}{(2\pi)^d}  \int_{\RR^d} \e^{\im \bq \cdot \bz} \hat{N}(\bq) [W_{\ve}(t,\bx,\bk-\bq) - W_{\ve}(t,\bx,\bk)] \,\rd \bq\,,
\end{multline}
where $\bmu \in \Lambda^*$, $\bp \in \mathcal{B}$, $ \bq \in \RR^d$.
Similar to the previous section, we introduce the fast variable $\bz = \frac{\bx}{\ve}$ in $W_{\ve}(t,\bx,\bk)$ and express the solution as $W_{\ve}(t,\bx,\bz,\bk)$. Then, \eqref{Wve} can be rewritten as follows:
\begin{multline}\label{Wve-full}
    \partial_t W_{\ve} + \bk \cdot \nabla_\bx W_{\ve} + \frac{1}{\ve} \bk \cdot \nabla_\bz W_{\ve} + \frac{\im \ve}{2} \Delta_\bx W_{\ve} + \im \nabla_\bx \cdot \nabla_\bz W_{\ve} + \frac{\im}{2 \ve} \Delta_\bz W_{\ve} \\[4pt]
    = \frac{1}{\im \ve} \sum_{\bmu \in \Lambda^*} \e^{\im \bmu \cdot \bz} \hat{V}(\bmu) \left[ W_{\ve}(t,\bx,\bz,\bk-\bmu) - W_{\ve}(t,\bx,\bz,\bk) \right]\\[4pt]
    + \frac{1}{\im \sqrt{\ve}} \frac{1}{(2\pi)^d} \int_{\RR^d} \hat{N}(\bq) [W_{\ve}(t,\bx,\bz,\bk-\bq)-W_{\ve}(t,\bx,\bz,\bk)] \,\rd \bq\,.
\end{multline}

Now we conduct the asymptotic analysis by substituting the following expansion into \eqref{Wve-full},
\begin{equation*}
    W_{\ve}(t,\bx,\bz,\bk) = W_0(t,\bx,\bz,\bk) + \sqrt{\ve} W_1(t,\bx,\bz,\bk) + \ve W_2(t,\bx,\bz,\bk) +....
\end{equation*}

Equating terms with like orders, we obtain the following equations. 

\textbf{(I) At $O(\frac{1}{\ve})$}, the equation reads:
\begin{equation*}
     \mathcal{L}[W_{0}](t,\bx,\bz,\bk) = 0\,.
\end{equation*}
Following the same reasoning as in the deterministic case, $W_0$ assumes the identical form as the expression in \eqref{W0}. Notably, in \eqref{W0}, $\sigma_{mn}(t,\bx,\bp)$ is initially defined for $\bp \in \mathcal B$. Nevertheless,  it is viable to extend this definition periodically, ensuring $\Lambda^*$-periodicity in $\bp$. This extension is admissible due to the $\Lambda^*$-periodicity of $Q_{mn}$ defined in \eqref{Qmn} with respect to $\bp$.

\textbf{(II) At $O(\frac{1}{\sqrt{\ve}})$}, we have 
\begin{equation}\label{O1/2}
    \mathcal{L}[W_1](t,\bx,\bz,\bk) =  \frac{1}{\im} \frac{1}{(2\pi)^d} \int_{\RR^d} \e^{\im \bq \cdot \bz} \hat{N}(\bq) [W_{0}(t,\bx,\bz,\bk-\bq) - W_{0}(t,\bx,\bz,\bk)] \,\rd \bq,
\end{equation}
from which we aim to derive an explicit formula for $W_1$. Since $W_1$ need not be periodic in $\bz$, it may not be expanded in $Q_{mn}$. Instead, we define: 
\begin{equation*}
    P_{mn} (\bz,\bmu, \bp, \bq) = \frac{1}{|\mathcal{C}|} \int_{\mathcal{C}} \e^{\im(\bp+\bmu) \cdot \by} \Psi_m(\bz-\by,\bp) \, \overline{\Psi_n(\bz,\bp+\bq)} \,\rd \by,
\end{equation*}
for $\bz \in \RR^d$ and $\bp, \bq \in \mathcal B$. Here $\Psi_m$ and $\Psi_n$ are the Bloch states defined in \eqref{FB}. Then, $P_{mn}$ are $\Lambda-$quasi-periodic in $\bz$, i.e.,
\begin{align*}
    P_{mn}(\bz + \bnu, \bmu, \bp, \bq) = P_{mn}(\bz, \bmu, \bp, \bq) \e^{- \im \bnu \cdot \bq}\,.
\end{align*}
Utilizing \eqref{25} and \eqref{26}, one can verify that $P_{mn}$ adheres to the following orthogonal property:
\begin{equation}\label{Ortho_P}
    \sum_{\bmu \in \Lambda^*} \frac{1}{|\mathcal{B}|} \int_{\RR^d}  P_{mn}(\bz,\bmu,\bp,\bq) \, \overline{P_{jl}(\bz,\bmu,\bp,\bq_0)} \,\rd \bz = \delta_{mj} \delta_{nl} \delta_{\text{per}}(\bq-\bq_0).
\end{equation}
Additionally, 
\begin{align*}
    \mathcal L [P_{mn}] = \im (E_m(\bp) - E_n(\bp+\bq)) P_{mn} (\bz, \bmu, \bp, \bq)\,.
\end{align*}

We seek $W_1(t,\bx,\bz,\bp+\bmu)$ in the following form:
\begin{equation}\label{W1}
    W_1(t,\bx,\bz,\bp+\bmu) = \sum_{m,n = 1}^2 \frac{1}{|\mathcal{B}|}  \int_{\mathcal{B}} \eta_{mn}(t,\bx,\bp,\bq) \, P_{mn}(\bz,\bmu,\bp,\bq) \,\rd \bq,
\end{equation}
for $\bz \in \RR^d$, $\bp\in \mathcal{B}$ and $\bmu \in \Lambda^*$, where the $\eta_{mn}$ are to be determined.

More specifically, by substituting $W_1$ from \eqref{W1} back into \eqref{O1/2} and multiplying both sides of \eqref{O1/2} by $\overline{P_{jl}(\bz,\bmu,\bp,\bq_0)}$, integrating over $\bz \in \RR^d$, and summing over $\bmu \in \Lambda^*$, we discover that the left-hand side of \eqref{O1/2} transforms into:
\begin{equation} \label{LHS}
\begin{split}
    \text{L.H.S} = & \left< \mathcal{L}[W_1](t,\bx,\bz,\bk) \,, P_{jl}(\bz,\bmu,\bp,\bq_0) \right>_{\mathcal{C},\Lambda^*} \\
    =& \sum_{\bmu \in \Lambda^*} \int_{\RR^d} \frac{1}{|\mathcal{B}|} \sum_{m,n=1}^2 \int_{\mathcal{B}}  \eta_{mn} \im [E_m(\bp)-E_n(\bp+\bq)] P_{mn}(\bz,\bmu,\bp,\bq) \overline{P_{jl}(\bz,\bmu,\bp,\bq_0)} \,\rd \bq \,\rd \bz \\
    =& \sum_{m,n=1}^2 \int_\mathcal{B}  \eta_{mn} \im [E_m(\bp)-E_n(\bp+\bq)] \delta_{mj} \delta_{nl} \delta_{\text{per}}(\bq-\bq_0) \,\rd \bq \\
    =&  \im  \eta_{jl} [E_j(\bp)-E_l(\bp+\bq_0)], \quad \bp, ~\bq_0 \in \mathcal B\,.
\end{split}
\end{equation}

The right-hand side of \eqref{O1/2} is
\begin{multline}\label{RHS}
    \text{R.H.S} = \frac{1}{\im}\frac{1}{(2\pi)^d} \sum_{\bmu \in \Lambda^*} \frac{1}{|\mathcal{B}|}\int_{\RR^d} \int_{\RR^d} \e^{\im \bq \cdot \bz} \hat{N}(\bq) [W_0(t, \bx, \bz,\bp+\bmu-\bq)\\ - W_0(t, \bx, \bz,\bp+\bmu)]  \overline{ P_{jl}(\bz,\bmu,\bp,\bq_0) } \,\rd \bz \,\rd \bq.
\end{multline}

Comparing \eqref{LHS} with \eqref{RHS}, we have that 
\begin{multline}\label{etajl}
    \eta_{jl}(t,\bx,\bp,\bq_0) = \frac{1}{(2\pi)^d} \sum_{\bmu \in \Lambda^*} \frac{1}{|\mathcal{B}|}\\[4pt]
    \int_{\RR^d} \int_{\RR^d} \e^{\im \bq \cdot \bz} \hat{N}(\bq) \frac{[W_0(t, \bx, \bz,\bp+\bmu-\bq) - W_0(t, \bx, \bz,\bp+\bmu)]}{E_l(\bp + \bq_0) - E_j(\bp) + \im \theta} \overline{ P_{jl}(\bz,\bmu,\bp,\bq_0) } \,\rd \bz \,\rd \bq,
\end{multline}
where $\theta > 0$ is introduced as a regularizing parameter and will be set to zero eventually. By substituting the expression for $W_0$ from \eqref{W0} into \eqref{etajl} and conducting a lengthy derivation, we arrive at, for $j,l \in \{1,2\}$,
\begin{equation}\label{etajl2}
\begin{split}
    \eta_{jl}(t,\bx,\bp,\bq_0) = & \frac{1}{(2\pi)^d} \sum_{\bmu \in \Lambda^*} \frac{\hat{N}(-\bmu-\bq_0)  \sum\limits_{m,n=1}^2 \sigma_{mn}(\bp+\bq_0)  \delta_{nl}} {E_l(\bp+\bq_0)-E_j(\bp) +\im \theta} A_{mj}(\bp+\bq_0+\bmu, \bp) \\
    & \hspace{-1.5cm} - \frac{1}{(2\pi)^d} \frac{1}{|\mathcal{B}|} \int_{\RR^d} \int_{\RR^d} \e^{\im \bq \cdot \bz} \frac{\hat{N}(\bq) \sum\limits_{m,n=1}^2 \sigma_{mn}(\bp) \delta_{mj}}{E_l(\bp+\bq_0) - E_j(\bp) + \im \theta}  \Psi_l(\bz,\bp+\bq_0) \overline{\Psi_n(\bz,\bp)} \,\rd \bz \,\rd \bq  \\[6pt]
    =:& \, (I) + (II)\,,
\end{split}
\end{equation}
where $A_{mj}(\bq,\bp)$ is defined by
\begin{equation*}
    A_{mj}(\bq,\bp) = \frac{1}{|\mathcal{C}|} \int_{\mathcal{C}} \e^{-\im(\bq-\bp) \cdot \by} \Psi_m(\by,\bq) \, \overline{\Psi_j(\by,\bp)} \,\rd \by.
\end{equation*}
The detailed derivation from \eqref{etajl} to \eqref{etajl2} can be found in Appendix \ref{appendix-B}.

\textbf{(III) At $O(1)$}, the equation reads:
\begin{multline}\label{O1-random}
    \frac{\partial W_0}{\partial t} + \bk \cdot \nabla_\bx W_0 + \im \nabla_\bz \cdot \nabla_\bx W_0 + \mathcal{L}[W_2] \\[1pt]
    = \frac{1}{\im} \frac{1}{(2\pi)^d} \int_{\RR^d} \e^{\im \bq \cdot \bz} \hat{N}(\bq) [W_1(t,\bx,\bz,\bk-\bq) - W_1(t,\bx,\bz,\bk)] \,\rd \bq - \frac{1}{\ve} \mathcal L [W_0]\,.
\end{multline}

To derive the system satisfied by $\sigma_{jl}$ with $j,l \in \{1,2\}$, it is necessary to multiply both sides of \eqref{O1-random} by $\overline{Q_{jl}(\bz,\bmu_1,\bp)}$, integrate over $\bz \in \RR^d$, and sum over $\bmu_1 \in \Lambda^*$ (Here the subscript of $\bmu$ is just added to distinguish different $\bmu$ that appear later on). The resulting equation is as follows. %for $\bk = \bp + \bmu_1$,

On the left-hand side, following a similar derivation as in the deterministic case, we have 
\begin{equation}\label{LHS-O1-R}
\begin{split}
    \text{L.H.S} = 
    \partial_t \sigma_{jl} + \sum_{n=1}^{2}  \nabla_\bx \sigma_{j n} \cdot \left<  (-\im \nabla_\bz) 
 \Psi_{l} \,, \Psi_n \right>_{\mathcal{C}} + \left< \mathcal{L}[W_2], \, Q_{jl} \right>_{\mathcal{C},\Lambda^*}.
\end{split}
\end{equation}
As elucidated similarly in \eqref{order1-sigma-full}, in what follows we will neglect the last term above.

On the right-hand side, we see that
\begin{equation*}
    \begin{split}
    \text{R.H.S} =& \frac{1}{\im} \frac{1}{(2\pi)^d} \frac{1}{|\mathcal{C}|} \sum_{\bmu_1 \in \Lambda^*} \int_{\mathcal{C}} \int_{\RR^d} \e^{\im \bq \cdot \bz} \hat{N}(\bq) W_1(t,\bx,\bz,\bp+\bmu_1-\bq) \, \overline{Q_{jl}(\bz,\bmu_1,\bp)} \,\rd \bq \,\rd \bz \\[6pt]
    &- \frac{1}{\im} \frac{1}{(2\pi)^d}  \frac{1}{|\mathcal{C}|} \sum_{\bmu_1 \in \Lambda^*} \int_{\mathcal{C}} \int_{\RR^d} \e^{\im \bq \cdot \bz} \hat{N}(\bq) W_1(t,\bx,\bz,\bp+\bmu_1) \, \overline{Q_{jl}(\bz,\bmu_1,\bp)} \,\rd \bq \,\rd \bz\\
    & - \frac{1}{\ve} \left<\mathcal{L}[W_0], Q_{jl} \right>_{\mathcal{C},\Lambda^*}\\[4pt]
    =& I_1 + I_2 + \frac{\im}{\ve} [E_l(\bp) - E_j(\bp)] \sigma_{jl} \, .
    \end{split}
\end{equation*}

\textbf{For $I_1$}, inserting the formula \eqref{W1} for $W_1$, we have
\begin{multline*}
    I_1 = \frac{1}{\im} \frac{1}{(2\pi)^d} \frac{1}{|\mathcal{C}|} \sum_{\bmu_1 \in \Lambda^*} \int_{\mathcal{C}} \int_{\RR^d} \e^{\im \bq \cdot \bz} \hat{N}(\bq) \sum_{m,n = 1}^2 \frac{1}{|\mathcal{B}|}  \int_{\mathcal{B}} \eta_{mn}(t,\bx,\bp+\bmu_1-\bq, \bq_0)\\
    \times P_{mn}(\bz,\bmu_1,\bp-\bq,\bq_0)  \,\overline{Q_{jl}(\bz,\bmu_1,\bp)} \,\rd \bq_0 \,\rd \bq \,\rd \bz\,,
\end{multline*}
which, upon substituting \eqref{etajl2} of $\eta_{mn}$, becomes 
\begin{equation*}
    \begin{split}
        I_1 
        =& \frac{1}{\im} \frac{1}{(2\pi)^d} \frac{1}{|\mathcal{C}|} \sum_{\bmu_1 \in \Lambda^*} \int_{\mathcal{C}} \int_{\RR^d} \e^{\im \bq \cdot \bz} \hat{N}(\bq) \sum\limits_{m,n = 1}^2 \frac{1}{|\mathcal{B}|}  \int_{\mathcal{B}} \\[6pt]
        &\times \frac{1}{(2\pi)^d} \sum_{\bmu_2 \in \Lambda^*} \frac{\hat{N}(-\bmu_2-\bq_0) \sum\limits_{m',n'=1}^2 \sigma_{m'n'}((\bp+\bmu_1-\bq)+\bq_0) \delta_{n'n} }{E_n((\bp+\bmu_1-\bq)+\bq_0)-E_m((\bp+\bmu_1-\bq)) +\im \theta} \\[6pt]
        & \qquad \qquad\qquad\qquad\qquad\qquad \times A_{m'm}((\bp+\bmu_1-\bq)+\bq_0+\bmu_2, (\bp+\bmu_1-\bq)) \\[6pt]
    &\times P_{mn}(\bz,\bmu_1,\bp-\bq,\bq_0) \, \overline{Q_{jl}(\bz,\bmu_1,\bp)} \,\rd \bq_0 \,\rd \bq \,\rd \bz\\[8pt]
    - & \frac{1}{\im} \frac{1}{(2\pi)^d} \frac{1}{|\mathcal{B}|} \frac{1}{|\mathcal{C}|} \sum_{\bmu_1 \in \Lambda^*} \int_{\mathcal{C}} \int_{\RR^d} \e^{\im \bq \cdot \bz} \hat{N}(\bq) \sum\limits_{m,n = 1}^2 \frac{1}{|\mathcal{B}|}  \int_{\mathcal{B}} \\[6pt]
    &\times \frac{1}{(2\pi)^d} \int_{\RR^d} \int_{\RR^d} \e^{\im \bq_1 \cdot \bz_1} \frac{\hat{N}(\bq_1) \sum\limits_{m',n'=1}^2 \sigma_{m'n'}((\bp+\bmu_1-\bq)) \delta_{mm'} }{E_n((\bp+\bmu_1-\bq)+\bq_0) - E_m((\bp+\bmu_1-\bq)) + \im \theta}  \\[6pt]
    & \qquad \qquad\qquad\qquad\quad  \times \Psi_n(\bz_1,(\bp+\bmu_1-\bq) + \bq_0) \, \overline{\Psi_{n'}(\bz_1,(\bp+\bmu_1-\bq))} \,\rd \bz_1 \,\rd \bq_1 \\[6pt]
    &\times P_{mn}(\bz,\bmu_1,\bp-\bq,\bq_0) \, \overline{Q_{jl}(\bz,\bmu_1,\bp)} \,\rd \bq_0 \,\rd \bq \,\rd \bz\\[8pt]
    =:& I_{11} - I_{12}\,.
    \end{split}
\end{equation*}

\textbf{For the term $I_{11}$}, taking the expectation and using the homogeneity relation \eqref{NNR} of the random process $N$, i.e.,
\begin{equation*}
   \expec [\hat{N}(-\bmu_2- \bq_0) \hat{N}(\bq)]  = (2\pi)^d \hat{R}(\bq) \, \delta(\bq-\bmu_2-\bq_0)\,.
\end{equation*}
Then upon integration over $\bq_0$, $I_{11}$ becomes
\begin{equation*}
    \begin{split}
        I_{11} =& \frac{1}{\im} \frac{1}{|\mathcal{C}|} \frac{1}{|\mathcal{B}|} \sum_{\bmu_1 \in \Lambda^*} \int_{\mathcal{C}} \int_{\RR^d} \e^{\im \bq \cdot \bz} \hat{R}(\bq) \sum_{m,n = 1}^2  \\[6pt]
        &\times \frac{1}{(2\pi)^d} \frac{\sum\limits_{m',n'=1}^2 \sigma_{m'n'}((\bp+\bmu_1-\bq)+(\bq-\bmu_2)) \delta_{n'n} }{E_n((\bp+\bmu_1-\bq)+(\bq-\bmu_2))-E_m((\bp+\bmu_1-\bq)) + \im \theta} \\[6pt]
        & \qquad \qquad\qquad\qquad \times A_{m'm}((\bp+\bmu_1-\bq)+(\bq-\bmu_2)+\bmu_2, (\bp+\bmu_1-\bq)) \\[6pt]
    &\times P_{mn}(\bz,\bmu_1,\bp-\bq,(\bq-\bmu_2))  \, \overline{Q_{jl}(\bz,\bmu_1,\bp)}  \,\rd \bq \,\rd \bz\,. \\
    \end{split}
\end{equation*}
Further, by taking into account the orthogonality of the Bloch eigenfunction in $P_{mn}$ and $\overline{Q_{jl}}$, $\Lambda^*$-periodicity of $\sigma_{m'n'}$, $E_m$, and $E_n$, and summing over $\bmu_1, \bmu_2 \in \Lambda^*$, $I_{11}$ can be simplified:
\begin{multline*}
        I_{11} = \sum_{m,n = 1}^2 \sum_{m',n'=1}^2 \frac{1}{\im} \frac{1}{(2\pi)^d} \frac{1}{|\mathcal{B}|} \sum_{\bmu \in \Lambda^*} \int_{\mathcal{B}} \hat{R}(\bq+\bmu)  \frac{ \sigma_{m'n'}(\bp) \delta_{n'n} \delta_{nl}}{E_n(\bp)-E_m(\bp-\bq) + \im \theta} \\[6pt]
         \times A_{m'm}(\bp, \bp-\bq-\bmu) \, \overline{A_{jm}(\bp, \bp-\bq-\bmu)} \,\rd \bq \, ,
\end{multline*}
where we have used decomposition for $\bp \in \RR^d$: $\bp = \Bar{\bp} + \Bar{\bmu} $ with $\Bar{\bp} \in \mathcal{B}$ and $\Bar{\bmu} \in \Lambda^*$, and subsequently renamed $\Bar{\bp}$ and $\Bar{\bmu}$ back to $\bp$ and $\bmu$.

%----------------------------------------------------------

\textbf{For the term $I_{12}$}, we take a similar calculation, i.e., using the relation \eqref{NNR} of the random process $N$,
\begin{equation*}
    \expec [ \hat{N}(\bq_1) \hat{N}(\bq) ] = (2\pi)^d \hat{R}(\bq) \, \delta(\bq+\bq_1)
\end{equation*}
and integrating over $\bq_1$, $I_{12}$ becomes
\begin{equation*}
    \begin{split}
    I_{12}& =  \frac{1}{\im}  \frac{1}{|\mathcal{C}|} \frac{1}{|\mathcal{B}|} \sum_{\bmu_1 \in \Lambda^*} \int_{\mathcal{C}} \int_{\RR^d} \e^{\im \bq \cdot \bz} R(\bq) \sum_{m,n = 1}^2 \frac{1}{|\mathcal{B}|}  \int_{\mathcal{B}} \\[6pt]
    &\times \frac{1}{(2\pi)^d} \int_{\RR^d}  \e^{-\im \bq \cdot \bz_1} \frac{ \sum\limits_{m',n'=1}^2 \sigma_{m'n'}(\bp+\bmu_1-\bq) \delta_{m'm} }{E_n((\bp+\bmu_1-\bq)+\bq_0) - E_m((\bp+\bmu_1-\bq)) + \im \theta}  \\[6pt]
    & \qquad \qquad\qquad\qquad\qquad\qquad \times \Psi_n(\bz_1,(\bp+\bmu_1-\bq)+\bq_0) \overline{\Psi_{n'}(\bz_1,(\bp+\bmu_1-\bq))} \,\rd \bz_1  \\[6pt]
    &\times P_{mn}(\bz,\bmu_1,\bp-\bq,\bq_0)  \,\overline{Q_{jl}(\bz,\bmu_1,\bp)} \,\rd \bq_0 \,\rd \bq \,\rd \bz \,, 
    \end{split}
\end{equation*}
and then, by using the orthogonal property of the Bloch function in $P_{mn}$ and $\overline{Q_{jl}}$:
\begin{equation*}
    \frac{1}{|\mathcal{B}|} \int_{\RR^d} \overline{\Psi_{n}(\bz,\bp-\bq+\bq_0) } \, \Psi_{l}(\bz,\bp) \,\rd \bz = \delta_{nl} \delta_{\text{per}}(-\bq+\bq_0) \, ,
\end{equation*}
we can take the integration over $\bq_0$, summation over $\bmu_1$, and consider $\Lambda^*$-periodicity of $\sigma_{m'n'}$, $E_m$ and $E_n$ to obtain 
\begin{equation*}
    \begin{split}
    I_{12}& =  \frac{1}{\im}  \frac{1}{|\mathcal{B}|} \int_{\RR^d} R(\bq) \sum_{m,n = 1}^2   \frac{ \sum\limits_{m',n'=1}^2 \sigma_{m'n'}(\bp-\bq) \delta_{mm'} \delta_{nl} }{E_n(\bp) - E_m(\bp-\bq) + \im \theta}\\[6pt]
    & \qquad \quad \times \frac{1}{(2\pi)^d} \int_{\RR^d}  \e^{-\im \bq \cdot \bz_1} \Psi_n(\bz_1,\bp) \overline{\Psi_{n'}(\bz_1,\bp-\bq)} \,\rd \bz_1 \overline{A_{jm}(\bp, \bp-\bq)} \,\rd \bq.
    \end{split}
\end{equation*}
Furthermore, by the definition of $A_{nn'}$, we can simplify $I_{12}$ as follows:
\begin{equation*}
    \begin{split}
    I_{12}& =  \sum_{m,n = 1}^2 \sum_{m',n'=1}^2 \frac{1}{\im} \frac{1}{(2\pi)^d} \frac{1}{|\mathcal{B}|} \sum_{\bmu \in \Lambda^*}  \int_{\mathcal{B}}  R(\bq+\bmu)    \frac{  \sigma_{m'n'}(\bp-\bq) \delta_{mm'} \delta_{nl} }{E_n(\bp) - E_m(\bp-\bq) + \im \theta}\\[6pt]
    &\times A_{nn'}(\bp, \bp-\bq-\bmu) \, \overline{A_{jm}(\bp, \bp-\bq-\bmu)} \,\rd \bq \, ,
    \end{split}
\end{equation*}
where we still take the decomposition $\bp \in \RR^d \mapsto \Bar{\bp} + \Bar{\bmu} $ with $\Bar{\bp} \in \mathcal{B}$ and $\Bar{\bmu} \in \Lambda^*$, and rename $\Bar{\bp}, \Bar{\bmu}$ back into $\bp,\bmu$.

Thus, combining the simplest form of $I_{11}$ and $I_{12}$, we obtain,
\begin{equation*}
\begin{split}
    I_1 =&\  I_{11} - I_{12} \\[6pt]
    =& \sum_{m,n = 1}^2 \sum_{m',n'=1}^2 \frac{1}{\im} \frac{1}{(2\pi)^d} \frac{1}{|\mathcal{B}|} \sum_{\bmu \in \Lambda^*} \int_{\mathcal{B}} \hat{R}(\bq+\bmu)   \frac{ \sigma_{m'n'}(\bp) \delta_{n'n} \delta_{nl}}{E_n(\bp)-E_m(\bp-\bq) + \im \theta} \\[6pt]
        & \qquad  \times A_{m'm}(\bp, \bp-\bq-\bmu) \,\overline{A_{jm}(\bp, \bp-\bq-\bmu)} \,\rd \bq\\[6pt]
    &-\sum_{m,n = 1}^2 \sum_{m',n'=1}^2 \frac{1}{\im} \frac{1}{(2\pi)^d} \frac{1}{|\mathcal{B}|} \sum_{\bmu \in \Lambda^*}  \int_{\mathcal{B}}  R(\bq+\bmu)    \frac{  \sigma_{m'n'}(\bp-\bq) \delta_{m'm} \delta_{nl} }{E_n(\bp) - E_m(\bp-\bq) + \im \theta}\\[6pt]
    &\qquad \times A_{nn'}(\bp,\bp-\bq-\bmu) \, \overline{A_{jm}(\bp, \bp-\bq-\bmu)} \,\rd \bq \\[6pt]
    =& \sum_{m,n = 1}^2 \sum_{m',n'=1}^2 \frac{1}{\im} \frac{1}{(2\pi)^d} \frac{1}{|\mathcal{B}|} \sum_{\bmu \in \Lambda^*} \int_{\mathcal{B}}    \frac{ \hat{R}(\bq+\bmu)}{E_n(\bp)-E_m(\bp-\bq) + \im \theta} \overline{A_{jm}(\bp, \bp-\bq-\bmu)}\\[6pt]
    & \times \Big[ \sigma_{m'n'}(\bp) \delta_{n'n} \delta_{nl}  A_{m'm}(\bp, \bp-\bq-\bmu) - \sigma_{m'n'}(\bp-\bq) \delta_{m'm} \delta_{nl} A_{nn'}(\bp,\bp-\bq-\bmu)  \Big] \,\rd \bq \,. 
\end{split}
\end{equation*}

Similar to \cite{BFPR1999JSP}, one can verify $I_2 = \overline{I_1}$. Therefore, by taking the limit $\theta \rightarrow 0$ and the change of variable $\bq \mapsto \bp - \bq$, the R.H.S finally becomes
\begin{equation}\label{RHS-O1-r}
\begin{split}
    \text{R.H.S} =& \sum_{m,n = 1}^2 \sum_{m',n'=1}^2  \frac{1}{(2\pi)^{d-1}}  \frac{1}{|\mathcal{B}|} \sum_{\bmu \in \Lambda^*} \int_{\mathcal{B}}    \hat{R}(\bp-\bq+\bmu) \delta(E_n(\bp)-E_m(\bq)) \overline{A_{jm}(\bp, \bq-\bmu)}\\[4pt]
    & \times \Big[\sigma_{m'n'}(\bq) \delta_{m'm} \delta_{nl} A_{nn'}(\bp,\bq-\bmu)- \sigma_{m'n'}(\bp) \delta_{n'n} \delta_{nl}  A_{m'm}(\bp, \bq-\bmu) \Big] \,\rd \bq \\[4pt]
    & + \frac{\im}{\ve} [E_l(\bp) - E_j(\bp)] \sigma_{jl} \\[4pt]
    =&\sum_{m,n = 1}^2 \sum_{m',n'=1}^2  \frac{1}{(2\pi)^{d-1}} \frac{1}{|\mathcal{B}|} \sum_{\bmu \in \Lambda^*} \int_{\mathcal{B}}   \hat{R}(\bp-\bq+\bmu) \delta(E_l(\bp)-E_m(\bq)) \overline{A_{jm}(\bp, \bq-\bmu)}\\[4pt]
    & \times \Big[ \sigma_{mn'}(\bq)   A_{ln'}(\bp,\bq-\bmu)  -\sigma_{m'l}(\bp)   A_{m'm}(\bp, \bq-\bmu) \Big] \,\rd \bq \\[4pt]
    & + \frac{\im}{\ve} [E_l(\bp) - E_j(\bp)] \sigma_{jl} \\[4pt]
    = &\sum_{m = 1}^2 \sum_{m'=1}^2  \frac{1}{(2\pi)^{d-1}} \frac{1}{|\mathcal{B}|} \sum_{\bmu \in \Lambda^*} \int_{\mathcal{B}}   \hat{R}(\bp-\bq+\bmu) \delta(E_l(\bp)-E_m(\bq)) \overline{A_{jm}(\bp, \bq-\bmu)}\\[6pt]
    & \times \Big[ \sigma_{m m'}(\bq)   A_{lm'}(\bp,\bq-\bmu) - \sigma_{m'l}(\bp)   A_{m'm}(\bp, \bq-\bmu)  \Big] \,\rd \bq \\[6pt]
    & +  \frac{\im}{\ve} [E_l(\bp) - E_j(\bp)] \sigma_{jl}, \quad j,l \in \{1,2\},
\end{split}
\end{equation}
where repeated indexes are removed in the last equality above.\\

% + \frac{1}{\sqrt{\ve}} \left< \mathcal{L}[W_1], \overline{Q_{jl}}\right>_{\bz,\bmu}  \\[4pt]
%& + \kq{ \frac{1}{\sqrt{\ve}} \left< \frac{1}{\im} \frac{1}{(2\pi)^d} \int_{\RR^d} \e^{\im \bq \cdot \bz} \hat{N}(\bq) [W_{0}(t,\bx,\bz,\bk-\bq) - W_{0}(t,\bx,\bz,\bk)] \,\rd \bq ,  \overline{Q_{jl}}\right>_{\bz,\bmu}}

%\kq{I think in Chai-Jin-Li \cite{CJL2013KRM}, the reason why they carried the lower order term $O(1/\ve)$ is that they actually didn't do the asymptotic analysis when there is band crossings, instead they just took the inner product to the equation of $W_{\ve}$, but in our case, I think excluding the lower-order terms will provide a more clean final version. } 

Thus, focusing on the $O(1)$ term and combining the L.H.S in \eqref{LHS-O1-R} and R.H.S in \eqref{RHS-O1-r}, we finally obtain the following coupled system:
\begin{equation} \label{eq:final_system}
\begin{split}
     &\partial_t \sigma_{jl}(t,\bx,\bp) + \sum_{n=1}^{2}  \nabla_\bx \sigma_{j n}(t,\bx,\bp) \cdot \left<  (-\im \nabla_\bz) \Psi_{l} \,, \Psi_n \right>_{\mathcal{C}}  \\
    = &\sum_{m = 1}^2 \sum_{m'=1}^2  \frac{1}{(2\pi)^{d-1}} \frac{1}{|\mathcal{B}|} \sum_{\bmu \in \Lambda^*} \int_{\mathcal{B}}   \hat{R}(\bp-\bq+\bmu) \delta(E_l(\bp)-E_m(\bq)) \overline{A_{jm}(\bp, \bq-\bmu)}\\[6pt]
    & \times \Big[ \sigma_{m m'}(\bq)   A_{lm'}(\bp,\bq-\bmu) - \sigma_{m'l}(\bp)   A_{m'm}(\bp, \bq-\bmu)  \Big] \,\rd \bq \\[6pt]
    & +  \frac{\im}{\ve} [E_l(\bp) - E_j(\bp)] \sigma_{jl}, \quad j,l \in \{1,2\}.
\end{split}
\end{equation}
%\aw{I propose we remove this part} 
%For the sake of completeness, we also explicitly present the dynamical system of $\sigma_{11}$, $\sigma_{12}$, $\sigma_{21}$ and $\sigma_{22}$ in Appendix \ref{appendix-C}.
Note that the discussion of Section \ref{sec:reduction} in the deterministic case also applies to the model \eqref{eq:final_system}. More specifically, if taking the transformation \eqref{eq:transformation}, we can obtain the coupled system for $\widetilde{\sigma}_{jl}$ by involving the phase factors as follows:
\begin{equation}\label{sigma_tilde_random}
    \begin{split}
    &\partial_t \widetilde{\sigma}_{jl}(t,\bx,\bp) + \sum_{n=1}^{2}  \nabla_\bx \widetilde{\sigma}_{j n}(t,\bx,\bp) \cdot \e^{- \frac{\im}{\ve}[E_l(\bp) - E_n(\bp)]t} \left<  (-\im \nabla_\bz)  \Psi_{l} \,, \Psi_n \right>_{\mathcal{C}} \\
    =& \sum_{m = 1}^2 \sum_{m' = 1}^2 \frac{1}{(2\pi)^{d-1}} \frac{1}{|\mathcal{B}|} \sum_{\bmu \in \Lambda^*} \int_{\mathcal{B}}   \hat{R}(\bp-\bq+\bmu) \delta(E_l(\bp)-E_m(\bq)) \overline{A_{jm}(\bp, \bq-\bmu)}\\[6pt]
    & \times \Big[ 
    \e^{-\frac{\im}{\ve}[E_m(\bq) - E_{m'}(\bq) + E_l(\bp) - E_j(\bp)] t}
    \widetilde{\sigma}_{mm'}(\bq)  A_{lm'}(\bp,\bq-\bmu)\\
    &\qquad\qquad\qquad\qquad\qquad\qquad - 
    \e^{-\frac{\im}{\ve}[E_{m'}(\bp) - E_{j}(\bp)] t}
    \widetilde{\sigma}_{m'l}(\bp)   A_{m'm}(\bp, \bq-\bmu) \Big] \,\rd \bq.
    \end{split}
\end{equation}

Similar calculations as in Section \ref{sec:reduction} then show that, for $\bp$ sufficiently far away from band crossings, the system \eqref{sigma_tilde_random} reduces to \cite[Eq.~(3.15)]{BFPR1999JSP} for the diagonal $\sigma$'s, $\sigma_{jj}$, $j \in \{1,2\}$, while the off-diagonal $\sigma$'s, $\sigma_{jl}$, $j \neq l$, can be neglected because if they are initially zero, they remain zero for all $t$. We provide these calculations in Appendix \ref{appendix-C}. In the following section we focus on the dynamics of the system \eqref{eq:final_system} close to band crossings in the particular case of graphene.
%\begin{remark}
    %It is noteworthy that our derived system \eqref{sigma_tilde_random} is generally valid for both circumstances when $\bp$ is ``away" ($|\bp-\bp_*|\gg O(\ve)$) and ``close" ($|\bp-\bp_*| \ll  o(\ve)$) to the band crossings.\\
    %Besides, notice that whenever $\bp$ is ``away" from the band crossings, the system \eqref{sigma_tilde_random} reduces to \cite[Eq.~(3.15)]{BFPR1999JSP} for the diagonal terms $\widetilde{\sigma}_{jl}, j=l \in \{1,2\}$.  
    %We provide detailed derivations of the explicit system for both diagonal and off-diagonal terms in Appendix \ref{appendix-C} when $\bp$ is ``away" from crossings. We will illustrate the situation when $\bp$ is ``close" to band crossings for a specific application with relevance to the physics of graphene in the following section. 
%\end{remark}

%%%%%%%%%%%%%%%%%%%%%%%%%%%%%%%%%%%%%%%%%%%%%%%%%%%%%%%%%%%%%%%%%%

\section{Application: effective dynamics of wave-packets in graphene with randomness}
\label{sec:application}

In this section we will specialize the model \eqref{eq:final_system} derived in the previous section to the particular case of the Schr\"{o}dinger operator with a honeycomb potential, modeling the dynamics of the wave-function of an electron in graphene.

\begin{figure}[h]
    \centering
    \includegraphics[width=0.65\textwidth]{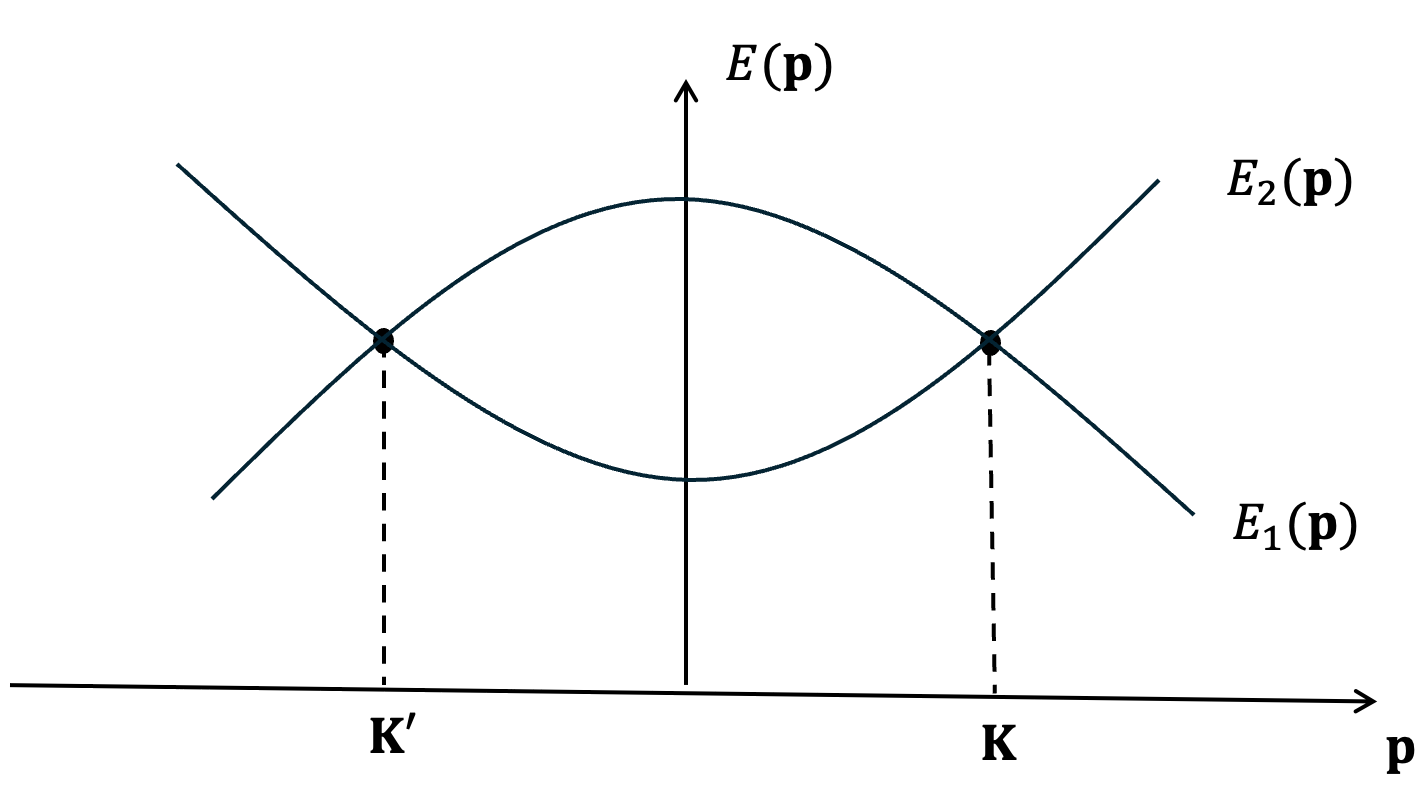}
    \caption{Schematic of energy band crossings for graphene.}
    \label{fig2}
\end{figure}

We start by reviewing the important features of the band structure of such operators, following \cite{FW2012JAMS, FW2014CMP, 2018BerkolaikoComech}. Generically, with respect to the magnitude of the potential \cite{FW2012JAMS}, we may assume that two bands are degenerate at the so-called ``Dirac points" in the Brillouin zone. These points are generally denoted by $\bK$ and $\bK' := - \bK$, see fig \ref{fig2}. Nearby to these points, the dispersion surface is conical, i.e., 
\begin{equation} \label{eq:conical}
    E_1(\bp) = - \lambda_\sharp |\bp - \bK| + o(|\bp - \bK|), \quad E_2(\bp) = \lambda_\sharp |\bp - \bK| + o(|\bp - \bK|), \quad \bp \rightarrow \bK,
\end{equation}
where $\lambda_\sharp$ is a positive constant known as the Fermi velocity \cite{NGPNG2009,FW2012JAMS}. In what follows we assume further that the bands $E_1, E_2$ are otherwise separated, as happens for tight-binding models \cite{NGPNG2009}. In this case \eqref{eq:conical} describes the shape of the bands as they intersect the Fermi level, which we take to be the zero of energy for convenience. To our knowledge, it is not known whether other eigenvalues can coincide with the Dirac energy for the full PDE model of graphene \cite{FW2012JAMS}.

%\kq{More physical interpretation of the derived system for graphene can be added here}

%The general derivation we have carried out in this work can be specialized to the case where the band degeneracy occurs only at two isolated points, i.e., ``Dirac points", as in graphene. It is interesting to consider how the derived system \eqref{eq:final_system} or \eqref{sigma_tilde_random} simplifies in this case. Suppose, for example, we consider initial conditions concentrated at the Dirac points, as in \cite{FW2014CMP}. Then, the $\sigma_{jl}$ will decay rapidly away from $\bp = \bK$ and $\bp = \bK'$, and we can obtain a simplified model by evaluating \eqref{eq:final_system} or \eqref{sigma_tilde_random} at these points. 
Recalling the discussion of different regimes in Section \ref{sec:reduction}, here we consider more precisely the case where $|\bp - \bK| = O(\ve)$ or $|\bp - \bK'| = O(\ve)$. This is the regime considered in \cite{FW2014CMP}, where $\ve$ corresponds to the wave-packet width in momentum space, and should be the one relevant for the physics of graphene for temperatures $T$ so that $k_B T \ll E$, where $k_B \approx 8.6 \times 10^{-5}$ $eV {K}^{-1}$ is Planck's constant and $E \approx 2.6$ $eV$ is the graphene bandwidth energy scale. Note that this holds even at room temperature $T \approx 300$ $K$. In the opposite case, where $|\bp - \bK| \gg \ve$ and $|\bp - \bK'| \gg \ve$, the system \eqref{eq:final_system} reduces to the result of Bal et al. \cite{BFPR1999JSP} as explained in Appendix \ref{appendix-C}.

Assuming $|\bp - \bK| = O(\ve)$ (the case $|\bp = \bK'| = O(\ve)$ is similar), we can obtain a closed system for $\sigma$'s with momenta near to the Dirac points by observing that, under our assumptions on the graphene band structure near to $0$ energy, the $\delta(E_l(\bp) - E_m(\bq))$ appearing in \eqref{eq:final_system} can be non-zero only for $|\bq - \bK| = O(\ve)$ or $|\bq - \bK'| = O(\ve)$. We can then replace \eqref{eq:final_system} by the simplified system
\begin{equation} \label{eq:reduced_graphene_0}
\begin{split}
     &\partial_t \sigma_{jl}(t,\bx, \bp) + \sum_{n=1}^{2}  \nabla_\bx \sigma_{j n}(t,\bx, \bp)  \cdot \left<  (-\im \nabla_\bz) 
 \Psi_{l}(\cdot, \bp) \,, \Psi_n(\cdot, \bp) \right>_{\mathcal{C}}  \\
    = &\sum_{m = 1}^2 \sum_{m'=1}^2  \frac{1}{(2\pi)^{d-1}} \frac{1}{|\mathcal{B}|} \sum_{\bmu \in \Lambda^*} \int_{\mathcal{B}_\ve(\bK,\bK')}   \hat{R}(\bp-\bq+\bmu) \delta(E_l(\bp)-E_m(\bq)) \overline{A_{jm}(\bp, \bq-\bmu)}\\[6pt]
    & \times \Big[ \sigma_{m m'}(\bq)   A_{lm'}(\bp,\bq-\bmu) - \sigma_{m'l}(\bp)   A_{m'm}(\bp, \bq-\bmu)  \Big] \,\rd \bq \\[6pt]
    & +  \frac{\im}{\ve} [E_l(\bp) - E_j(\bp)] \sigma_{jl},
    %= &\sum_{m,n = 1}^2 \sum_{m',n'=1}^2  \frac{1}{(2\pi)^{d-1}}  \sum_{\bmu \in \Lambda^*} \sum_{\bq= \bK, \bK'}  \hat{R}(\bK-\bq+\bmu)  \overline{A_{jm}(\bK, \bq-\bmu)}\\[6pt]
    %& \times \Big[ \sigma_{m'l}(\bK)   A_{m'm}(\bK, \bq-\bmu) - \sigma_{mn'}(\bq)   A_{ln'}(\bK,\bq-\bmu)  \Big],
\end{split}
\end{equation}
where $\mathcal{B}_\ve(\bK,\bK')$ denotes the union of balls with radius proportional to $\ve$ centered at $\bK$ and $\bK'$. Note that the system \eqref{eq:reduced_graphene_0} couples only $\sigma$'s with momenta near to $\bK$ and $\bK'$, and that the final relaxation term in \eqref{eq:reduced_graphene_0} is $O(1)$ as $\ve \downarrow 0$ by our assumptions on $\bp$ and \eqref{eq:conical}. We can interpret equation \eqref{eq:reduced_graphene_0} as describing the evolution of wave-packets concentrated at Dirac points $\bK$ and $\bK'$ with momentum width $\ve$ coupled to each other by the random potential.

%where we assume that $E_l(\bK) = E_m(\bq) \iff \bq = \bK$ or $\bK'$ for $1 \leq l, m \leq 2$. Note that the $\Leftarrow$ implication is an assumption here: although it certainly holds for the nearest-neighbor tight-binding model of graphene \cite{NGPNG2009}, 

Further simplifications are possible if we assume that $|\bp - \bK| = o(\ve)$ or $|\bp - \bK'| = o(\ve)$. In this case, after dropping higher order terms, we can obtain a closed system for $\sigma_{jl}(t,\bx,\bp)$ for $\bp \in \{\bK,\bK'\}$ as 
    \begin{equation} \label{eq:reduced_graphene_minus1}
    \begin{split}
     &\partial_t \sigma_{jl}(t,\bx, \bK) + \sum_{n=1}^{2}  \nabla_\bx \sigma_{j n}(t,\bx, \bK)  \cdot \left<  (-\im \nabla_\bz) 
 \Psi_{l}(\cdot, \bK) \,, \Psi_n(\cdot, \bK) \right>_{\mathcal{C}}  \\
    %= &\sum_{m=1}^2 \sum_{m'=1}^2  \frac{1}{(2\pi)^{d-1}} \frac{1}{|\mathcal{B}|} \sum_{\bmu \in \Lambda^*} \int_{\mathcal{B}}   \hat{R}(\bK-\bq+\bmu) \delta(E_l(\bK)-E_m(\bq)) \overline{A_{jm}(\bK, \bq-\bmu)}\\[6pt]
    %& \times \Big[ \sigma_{mm'}(\bq)   A_{lm'}(\bK,\bq-\bmu) - \sigma_{m'l}(\bK)   A_{m'm}(\bK, \bq-\bmu) \Big] \,\rd \bq \\[6pt]
    = &\sum_{m = 1}^2 \sum_{m'=1}^2  \frac{1}{(2\pi)^{d-1}} \frac{1}{|\mathcal{B}|} \sum_{\bmu \in \Lambda^*} \sum_{\bq= \bK, \bK'}  \hat{R}(\bK-\bq+\bmu)  \overline{A_{jm}(\bK, \bq-\bmu)}\\[6pt]
    & \times \Big[ \sigma_{mm'}(\bq)   A_{lm'}(\bK,\bq-\bmu) - \sigma_{m'l}(\bK)   A_{m'm}(\bK, \bq-\bmu) \Big],
    \end{split}
    \end{equation}
    and
    \begin{equation} \label{eq:reduced_graphene_minus2}
    \begin{split}
     &\partial_t \sigma_{jl}(t,\bx, \bK') + \sum_{n=1}^{2}  \nabla_\bx \sigma_{j n}(t,\bx, \bK')  \cdot \left<  (-\im \nabla_\bz) 
 \Psi_{l}(\cdot, \bK') \,, \Psi_n(\cdot, \bK') \right>_{\mathcal{C}}  \\
    = &\sum_{m= 1}^2 \sum_{m'=1}^2  \frac{1}{(2\pi)^{d-1}} \frac{1}{|\mathcal{B}|} \sum_{\bmu \in \Lambda^*} \sum_{\bq= \bK, \bK'}  \hat{R}(\bK'-\bq+\bmu)  \overline{A_{jm}(\bK', \bq-\bmu)}\\[6pt]
    & \times \Big[ \sigma_{mm'}(\bq) A_{lm'}(\bK',\bq-\bmu) - \sigma_{m'l}(\bK') A_{m'm}(\bK', \bq-\bmu) \Big].
    \end{split}
    \end{equation}
Note that the off-diagonal $\sigma$'s cannot be ignored in either of the systems \eqref{eq:reduced_graphene_0} or \eqref{eq:reduced_graphene_minus1}-\eqref{eq:reduced_graphene_minus2}.

To our knowledge, the models \eqref{eq:reduced_graphene_0} or \eqref{eq:reduced_graphene_minus1}-\eqref{eq:reduced_graphene_minus2} are original to the present work. They are straightforward to interpret: in the presence of a weak random potential as in \eqref{random_system}, wave-packets concentrated at the Dirac points in graphene no longer satisfy independent dynamics but become coupled. A similar model where wave-packets propagating along domain walls in modulated graphene-like structures can become ``valley-coupled" through a random perturbation was recently introduced in \cite{BCMQ2023}. Note that the shape of the eigenvalue bands at the crossing, which is linear in graphene ($r = 1$ in \eqref{eq:lin_cross}), determines the regime where the reduced models we derive are valid. In the case of a crossing where $r \neq 1$, the regime would have to be modified from $|\bp - \bK| = O(\ve)$ to $|\bp - \bK| = O(\ve^{1/r})$.

\section{Conclusions}
\label{sec:conclusions}

In this paper, we investigate the semi-classical limit of the Schr\"{o}dinger equation featuring general periodic potentials across arbitrary dimensions. Our focus lies on situations where energy bands exhibit crossings, a characteristic particularly significant for materials such as graphene. In the absence of randomness, we develop a coupled Liouville system with a relaxation-type source term, capturing the interplay between energy bands. Conversely, when introducing random perturbations, we establish a coupled radiative transport system. Here, an additional collision-like term elucidates interactions between distinct wave vectors sharing the same energy. As a special case, we examine the potential of a honeycomb structure. Our newly derived system unveils that for wave packets concentrated at the Dirac points in graphene, they no longer exhibit independent dynamics but rather become coupled.

%%%%%%%%%%%%%%%%%%%%%%%%%%%%%%%%%%%%%%%%%%%%%%%%%%%%%%%%%%%%%%%%%%%%%

\appendix

\section{Derivation of orthogonality relation}
\label{appendix-A}

In this section, we present the complete derivation of the orthogonality \eqref{Qmnjl}. For $\bp \in \mathcal{B}$, we have
\begin{equation}\label{der_Qmnjl_1}
    \begin{split}
        & \left<  Q_{mn}(\cdot, \cdot, \bp) \,, Q_{jl}(\cdot, \cdot, \bp) \right>_{\mathcal{C},\Lambda^*} \\[4pt]
        =& \ \sum_{\bmu \in \Lambda^*} \frac{1}{|\mathcal{C}|} \int_{\mathcal{C}} Q_{mn}(\bz, \bmu, \bp) \,\overline{Q_{jl}(\bz, \bmu, \bp)}  \,\rd \bz \\[4pt]
        =& \  \sum_{\bmu \in \Lambda^*} \int_\mathcal{C} \frac{1}{|\mathcal{C}|} \left[ \int_\mathcal{C} \frac{1}{|\mathcal{C}|} \e^{\im(\bp+\bmu) \cdot \by} \Psi_m(\bz-\by,\bp) \overline{\Psi_n(\bz,\bp)} \,\rd \by \right] \\
    & \qquad \times \left[ \int_\mathcal{C} \frac{1}{|\mathcal{C}|} \e^{-\im(\bp+\bmu) \cdot \by'} \overline{\Psi_{j}(\bz-\by',\bp)} \Psi_{l}(\bz,\bp) \,\rd \by' \right] \,\rd \bz \\[4pt]
    =& \ \sum_{\bmu \in \Lambda^*} \int_\mathcal{C} \frac{1}{|\mathcal{C}|} \left[ \int_{\mathcal{C}'} \frac{1}{|\mathcal{C}|}  \e^{\im(\bp+\bmu) \cdot (\bz-\by)} \Psi_m(\by,\bp) \overline{\Psi_n(\bz,\bp)} \,\rd \by \right] \\
    & \qquad \times \left[ \int_{\mathcal{C}'} \frac{1}{|\mathcal{C}|} \e^{-\im(\bp+\bmu) \cdot (\bz-\by')} \overline{\Psi_{j}(\by',\bp)} \Psi_{l}(\bz,\bp) \,\rd \by' \right] \,\rd \bz
    \end{split}
    \end{equation}
    where we applied the change of variable 
\begin{equation*}
    \by \mapsto (\bz-\by) \qquad  \by' \mapsto (\bz-\by')
\end{equation*}
with the change of domain $ \mathcal{C} \mapsto \mathcal{C}' $ in the last equality above. By further integrating over $\bz$ and using the relation
\begin{equation*}
    \frac{1}{|\mathcal{C}|} \int_{\mathcal{C}}  \overline{\Psi_n(\bz,\bp)} \, \Psi_{l}(\bz,\bp) \,\rd \bz = \delta_{ln}
\end{equation*}
\eqref{der_Qmnjl_1} becomes
\begin{equation}
    \begin{split}
        & \delta_{ln} \frac{1}{|\mathcal{C}|} \sum_{\bmu \in \Lambda^*}   \int_{\mathcal{C}'} \int_{\mathcal{C}'} \frac{1}{|\mathcal{C}|} \e^{\im(\bp+\bmu)\cdot (\by'-\by)} \Psi_m(\by,\bp) \, \overline{\Psi_{j}(\by',\bp)} \,\rd \by  \,\rd \by'\\
        =& \delta_{ln}   \int_{\mathcal{C}'} \int_{\mathcal{C}'} \frac{1}{|\mathcal{C}|}  \sum_{\bnu \in \Lambda}\delta((\by'-\by)-\bnu) \e^{\im \bp (\by'-\by)} \Psi_m(\by,\bp) \, \overline{\Psi_{j}(\by', \bp)} \,\rd \by  \,\rd \by' \\
        =& \delta_{ln}   \int_{\mathcal{C}'} \int_{\mathcal{C}'} \frac{1}{|\mathcal{C}|}  \sum_{\bnu \in \Lambda}\delta((\by'-\by)-\bnu) \e^{\im \bp (\by'-\by)} \Psi_m(\by,\bp) \,  \overline{\e^{\im \bp \cdot \bnu} \Psi_{j}(\by'-\bnu, \bp)} \,\rd \by  \,\rd \by' \\
        =& \delta_{ln}   \int_{\mathcal{C}'} \int_{\mathcal{C}'} \frac{1}{|\mathcal{C}|}  \sum_{\bnu \in \Lambda}\delta((\by'-\by)-\bnu) \e^{\im \bp (\by'-\bnu-\by)} \Psi_m(\by,\bp) \, \overline{\Psi_{j}(\by'-\bnu, \bp)} \,\rd \by  \,\rd \by' \\
        =& \delta_{ln}    \int_{\mathcal{C}'} \frac{1}{|\mathcal{C}|}  \Psi_m(\by,\bp) \, \overline{\Psi_{j}(\by,\bp)} \,\rd \by   \\ 
        = & \delta_{ln} \delta_{jm}
    \end{split}
\end{equation}
where we use the following identity \cite[Appendix A]{vanderbilt2018berry} in the first equality above
\begin{equation}\label{26}
     \frac{1}{|\mathcal{C}|} \sum_{\bmu \in \Lambda^*} \e^{\im\bmu\cdot \bz} = \sum_{\bnu \in \Lambda} \delta(\bz-\bnu) \Rightarrow \frac{1}{|\mathcal{C}|} \sum_{\bmu \in \Lambda^*} \e^{\im \bmu \cdot (\by'-\by)} = \sum_{\bnu \in \Lambda} \delta(\by'-\by-\bnu) \,,
\end{equation}
and the periodicity of Bloch eigenfunction $\eqref{FB}_2$ in the second equality. 

By involving the integration by parts and product rule, it then yields the \eqref{kQmnjl} and \eqref{gQmnjl} by following similar calculations, respectively.

%-----------------------------------------------

\section{Derivation of $\eta_{jl}$}
\label{appendix-B}

In this section, we provide the specific calculation process of $\eta_{jl}$, i.e., from \eqref{etajl} to \eqref{etajl2}. To clarify, we first present the derivation of $(I)$ in \eqref{etajl2},
\begin{equation*}
    \begin{split}
        (I) = &\frac{1}{(2\pi)^d} \frac{1}{|\mathcal{B}|} \sum_{\bmu \in \Lambda^*} \int_{\RR^d} \int_{\RR^d} \e^{\im \bq \cdot \bz} \hat{N}(\bq) W_0(\bz,\bp+\bmu-\bq) \, \overline{ P_{jl}(\bz,\bmu,\bp,\bq_0) } \,\rd \bz \,\rd \bq \\[4pt]
        = & \frac{1}{(2\pi)^d} \frac{1}{|\mathcal{B}|} \sum_{\bmu \in \Lambda^*} \int_{\RR^d} \int_{\RR^d} \e^{\im \bq \cdot \bz} \hat{N}(\bq) \left[\sum_{m,n=1}^2 \sigma_{mn}(t,\bx,\bp-\bq) Q_{mn}(\bz,\bmu,\bp-\bq) \right] \\[4pt]
        &  \frac{1}{|\mathcal{C}|} \int_{\mathcal{C}} \e^{-\im(\bp+\bmu) \cdot \by_2} \overline{\Psi_j(\bz-\by_2,\bp)} \, \Psi_l(\bz,\bp+\bq_0) \,\rd \by_2 \,\rd \bz \,\rd \bq \\[4pt]
        = & \frac{1}{(2\pi)^d} \frac{1}{|\mathcal{B}|} \sum_{\bmu \in \Lambda^*} \int_{\RR^d} \int_{\RR^d} \e^{\im \bq \cdot \bz} \hat{N}(\bq) \\
        & \times \left[\sum_{m,n=1}^2 \sigma_{mn}(t,\bx,\bp-\bq) \frac{1}{|\mathcal{C}|} \int_{\mathcal{C}} \e^{\im(\bp+\bmu-\bq) \cdot \by_1} \Psi_m(\bz-\by_1,\bp-\bq) \, \overline{\Psi_n(\bz,\bp-\bq)} \,\rd \by_1 \right] \\[4pt]
        & \times \frac{1}{|\mathcal{C}|} \int_{\mathcal{C}} \e^{-\im(\bp+\bmu) \cdot \by_2} \overline{\Psi_j(\bz-\by_2,\bp)} \, \Psi_l(\bz,\bp+\bq_0) \,\rd \by_2 \,\rd \bz \,\rd \bq \,.
    \end{split}
\end{equation*}
By using the following change of variable :
\begin{equation*}
    \by_1 \mapsto \bz-\by_1 \quad \quad \by_2 \mapsto \bz-\by_2,
\end{equation*}
it yields that
\begin{equation*}
    \begin{split}
        (I) = & \frac{1}{(2\pi)^d} \frac{1}{|\mathcal{B}|} \sum_{\bmu \in \Lambda^*} \int_{\RR^d} \int_{\RR^d}  \hat{N}(\bq) \sum_{m,n=1}^2 \sigma_{mn}(t,\bx,\bp-\bq) \\[4pt]
        &  \frac{1}{|\mathcal{C}|^2} \int_{\mathcal{C}'\times \mathcal{C}'} \e^{\im(\bp+\bmu) \cdot (
        \by_2-\by_1)} \e^{\im \bq \cdot \by_1} \Psi_m(\by_1,\bp-\bq) \, \overline{\Psi_j(\by_2,\bp)} \\
        & \qquad \qquad \qquad \qquad \qquad  \times \Psi_l(\bz,\bp+\bq_0) \, \overline{\Psi_n(\bz,\bp-\bq)} \,\rd \by_1 \,\rd \by_2 \,\rd \bz \,\rd \bq\\[4pt]
        = & \frac{1}{(2\pi)^d} \sum_{\bmu \in \Lambda^*}  \int_{\RR^d}  \hat{N}(\bq) \sum_{m,n=1}^2 \sigma_{mn}(t,\bx,\bp-\bq)  \delta_{nl} \delta_{\text{per}}(-\bq-\bq_0)   \\
        &  \frac{1}{|\mathcal{C}|^2} \int_{\mathcal{C}'\times \mathcal{C}'} \e^{\im(\bp+\bmu) \cdot (\by_2-\by_1)} \e^{\im \bq \cdot \by_1} \Psi_m(\by_1,\bp-\bq) \, \overline{\Psi_j(\by_2,\bp)}  \,\rd \by_1 \,\rd \by_2  \,\rd \bq\,,
    \end{split}
\end{equation*}
where the orthogonality \eqref{25} is used in the last equality.

Taking another change of variable 
\begin{equation*}
    \bq \mapsto \Bar{\bq} - \Bar{\bmu}, \quad  \text{with} \quad \bq \in \RR^d, \quad \Bar{\bq} \in \mathcal{B}, \quad \Bar{\bmu} \in \Lambda^*\,,
\end{equation*}
% we have $\int_{\RR^d} F(\bq) \,\rd \bq = \sum_{\Bar{\bmu} \in \Lambda^*} \int_{\mathcal{B}}  F(\Bar{\bq} - \Bar{\bmu}) \,\rd \Bar{\bq}\,,$
% and then
\begin{equation*}
\begin{split}
      (I) = & \frac{1}{(2\pi)^d} \sum_{\bmu \in \Lambda^*} \sum_{\Bar{\bmu} \in \Lambda^*}  \int_{\mathcal{B}}  \hat{N}(\Bar{\bq} - \Bar{\bmu}) \sum_{m,n=1}^2 \sigma_{mn}(t,\bx,\bp-(\Bar{\bq} - \Bar{\bmu})) \delta_{nl}    \\[4pt]
        &  \times \frac{1}{|\mathcal{C}|^2} \int_{\mathcal{C}'\times \mathcal{C}'} \e^{\im(\bp+\bmu)\cdot (\by_2-\by_1)} \e^{\im (\Bar{\bq} - \Bar{\bmu}) \cdot \by_1}\Psi_m(\by_1,\bp-(\Bar{\bq} - \Bar{\bmu})) \overline{\Psi_j(\by_2,\bp)}\\[4pt]
        & \times \delta_{\text{per}}(-(\Bar{\bq} - \Bar{\bmu})-\bq_0) \,\rd \by_1 \,\rd \by_2  \,\rd \Bar{\bq}\\[4pt]
        = & \frac{1}{(2\pi)^d} \sum_{\bmu \in \Lambda^*} \sum_{\Bar{\bmu} \in \Lambda^*}  \int_{\mathcal{B}}  \hat{N}(\Bar{\bq} - \Bar{\bmu}) \sum_{m,n=1}^2 \sigma_{mn}(t,\bx,\bp-\Bar{\bq} ) \delta_{nl}    \\[4pt]
        &  \frac{1}{|\mathcal{C}|^2} \int_{\mathcal{C}' \times \mathcal{C}'} \e^{\im(\bp+\bmu)\cdot (\by_2-\by_1)} \e^{\im (\Bar{\bq} - \Bar{\bmu}) \cdot \by_1}\Psi_m(\by_1,\bp-(\Bar{\bq} - \Bar{\bmu})) \, \overline{\Psi_j(\by_2,\bp)} \\[4pt]
        & \times \delta_{\text{per}}(-\Bar{\bq} - \bq_0) \,\rd \by_1 \,\rd \by_2  \,\rd \Bar{\bq}\\
        = & \frac{1}{(2\pi)^d} \sum_{\bmu \in \Lambda^*} \sum_{\Bar{\bmu} \in \Lambda^*}  \hat{N}(-\bq_0 - \Bar{\bmu}) \sum_{m,n=1}^2 \sigma_{mn}(t,\bx,\bp+\bq_0 ) \delta_{nl}    \\[4pt]
        &  \frac{1}{|\mathcal{C}|^2} \int_{\mathcal{C}'\times \mathcal{C}'} \e^{\im(\bp+\bmu)\cdot (\by_2-\by_1)} \e^{\im (-\bq_0 - \Bar{\bmu}) \cdot \by_1} \Psi_m(\by_1,\bp-(-\bq_0 - \Bar{\bmu})) \, \overline{\Psi_j(\by_2,\bp)}  \,\rd \by_1 \,\rd \by_2\\
        & \frac{1}{(2\pi)^d}  \sum_{\Bar{\bmu} \in \Lambda^*}  \hat{N}(-\bq_0 - \Bar{\bmu}) \sum_{m,n=1}^2 \sigma_{mn}(t,\bx,\bp+\bq_0 ) \delta_{nl}    \\[4pt]
        &  \times \frac{1}{|\mathcal{C}|} \int_{\mathcal{C}' \times \mathcal{C}'} \e^{\im \bp \cdot (\by_2-\by_1)} \e^{\im (-\bq_0 - \Bar{\bmu}) \cdot \by_1} \Psi_m(\by_1,\bp-(-\bq_0 - \Bar{\bmu})) \\[4pt]
        & \times \overline{\Psi_j(\by_2,\bp)} \sum_{\bnu \in \Lambda} \delta(\by_2-\by_1-\bnu)  \,\rd \by_1 \,\rd \by_2 \\[4pt]
        = & \frac{1}{(2\pi)^d}  \sum_{\Bar{\bmu} \in \Lambda^*} \hat{N}(-\bq_0 - \Bar{\bmu}) \sum_{m,n=1}^2 \sigma_{mn}(t,\bx,\bp+\bq_0 ) \delta_{nl}    \\[4pt]
        &  \times \frac{1}{|\mathcal{C}|} \int_{\mathcal{C}' \times \mathcal{C}'} \e^{\im \bp \cdot (\by_2- \bnu-\by_1)} \e^{\im (-\bq_0 - \Bar{\bmu}) \cdot \by_1} \Psi_m(\by_1,\bp-(-\bq_0 - \Bar{\bmu})) \\[4pt]
        & \times \overline{\Psi_j(\by_2 - \bnu,\bp)} \sum_{\bnu \in \Lambda} \delta(\by_2-\by_1-\bnu)  \,\rd \by_1 \,\rd \by_2 \\[4pt]
        = & \frac{1}{(2\pi)^d}  \sum_{\Bar{\bmu} \in \Lambda^*}  \hat{N}(-\bq_0 - \Bar{\bmu}) \sum_{m,n=1}^2 \sigma_{mn}(t,\bx,\bp+\bq_0 ) \delta_{nl}  \\[4pt]
        & \times  \frac{1}{|\mathcal{C}|} \int_{\mathcal{C}'}  \e^{-\im (\bq_0 + \Bar{\bmu}) \cdot \by_1} \Psi_m(\by_1,\bp + \bq_0 + \Bar{\bmu}) \overline{\Psi_j(\by_1 ,\bp)}  \,\rd \by_1  \\[4pt]
        = & \frac{1}{(2\pi)^d}  \sum_{\bmu \in \Lambda^*}  \hat{N}(-\bq_0 - \bmu) \sum_{m,n=1}^2 \sigma_{mn}(t,\bx,\bp+\bq_0 ) \delta_{nl}  \\[4pt]
        &  \times \underbrace{\frac{1}{|\mathcal{C}|} \int_{\mathcal{C}'}  \e^{-\im (\bq_0 + \bmu) \cdot \by} \Psi_m(\by,\bp + \bq_0 + \bmu) \, \overline{\Psi_j(\by,\bp)}  \,\rd \by }_{:=A_{mj}(\bp+\bq_0+\bmu,\bp)}
\end{split}
\end{equation*}
where we consider the $\Lambda^*$-periodic $\sigma_{mn}$ in the second equality above and re-name $\by_1 \mapsto \by$ and $\Bar{\bmu} \mapsto \bmu$ in the last equality above.

To derive the term $(II)$ of \eqref{etajl2}, we start with 
\begin{equation*}
    \begin{split}
        (II)= &\frac{1}{(2\pi)^d} \frac{1}{|\mathcal{B}|} \sum_{\bmu \in \Lambda^*} \int_{\RR^d} \int_{\RR^d} \e^{\im \bq \cdot \bz} \hat{N}(\bq) W_0(\bz,\bp+\bmu) \, \overline{ P_{jl}(\bz,\bmu,\bp,\bq_0) } \,\rd \bz \,\rd \bq \\
        = & \frac{1}{(2\pi)^d} \frac{1}{|\mathcal{B}|} \sum_{\bmu \in \Lambda^*} \int_{\RR^d} \int_{\RR^d} \e^{\im \bq \cdot \bz} \hat{N}(\bq) \left[\sum_{m,n=1}^2 \sigma_{mn}(t,\bx,\bp) Q_{mn}(\bz,\bmu,\bp) \right] \\
        & \times  \frac{1}{|\mathcal{C}|} \int_{\mathcal{C}} \e^{-\im(\bp+\bmu) \cdot \by_2} \overline{\Psi_j(\bz-\by_2,\bp)} \, \Psi_l(\bz,\bp+\bq_0) \,\rd \by_2 \,\rd \bz \,\rd \bq \\[4pt]
        = & \frac{1}{(2\pi)^d} \frac{1}{|\mathcal{B}|} \sum_{\bmu \in \Lambda^*} \int_{\RR^d} \int_{\RR^d} \e^{\im \bq \cdot \bz} \hat{N}(\bq) \\
        &\times \left[\sum_{m,n=1}^2 \sigma_{mn}(t,\bx,\bp) \frac{1}{|\mathcal{C}|} \int_{\mathcal{C}} \e^{\im(\bp+\bmu) \cdot \by_1} \Psi_m(\bz-\by_1,\bp)  \, \overline{\Psi_n(\bz,\bp)} \,\rd \by_1 \right] \\
        & \times  \frac{1}{|\mathcal{C}|} \int_{\mathcal{C}} \e^{-\im(\bp+\bmu) \cdot \by_2} \overline{\Psi_j(\bz-\by_2,\bp)} \, \Psi_l(\bz,\bp+\bq_0) \,\rd \by_2 \,\rd \bz \,\rd \bq\\[4pt]
        =& \frac{1}{(2\pi)^d} \frac{1}{|\mathcal{B}|} \sum_{\bmu \in \Lambda^*} \int_{\RR^d} \int_{\RR^d} \e^{\im \bq \cdot \bz} \hat{N}(\bq) \sum_{m,n=1}^2 \sigma_{mn}(t,\bx,\bp)    \\[4pt]
        & \times  \frac{1}{|\mathcal{C}|^2} \int_{\mathcal{C}' \times \mathcal{C}'} \e^{\im(\bp+\bmu) \cdot (\by_2-\by_1)} \Psi_m(\by_1,\bp) \, \overline{\Psi_j(\by_2,\bp)} \, \Psi_l(\bz,\bp+\bq_0) \, \overline{\Psi_n(\bz,\bp)} \,\rd \by_1 \,\rd \by_2 \,\rd \bz \,\rd \bq\\[4pt]
        =& \frac{1}{(2\pi)^d} \frac{1}{|\mathcal{B}|} \int_{\RR^d} \int_{\RR^d} \e^{\im \bq \cdot \bz} \hat{N}(\bq) \sum_{m,n=1}^2 \sigma_{mn}(t,\bx,\bp)   \Psi_l(\bz,\bp+\bq_0) \, \overline{\Psi_n(\bz,\bp)} \\
        &  \times \frac{1}{|\mathcal{C}|} \int_{\mathcal{C}' \times \mathcal{C}'} \e^{\im \bp \cdot (\by_2-\by_1)} \sum_{\bnu \in \Lambda} \delta(\by_2-\by_1-\bnu) \Psi_m(\by_1,\bp) \, \overline{\Psi_j(\by_2,\bp)} \,\rd \by_1 \,\rd \by_2 \,\rd \bz \,\rd \bq\\[4pt]
        =& \frac{1}{(2\pi)^d} \frac{1}{|\mathcal{B}|} \int_{\RR^d} \int_{\RR^d} \e^{\im \bq \cdot \bz} \hat{N}(\bq) \sum_{m,n=1}^2 \sigma_{mn}(t,\bx,\bp)  \Psi_l(\bz,\bp+\bq_0) \,  \overline{\Psi_n(\bz,\bp)}  \\
        & \times \frac{1}{|\mathcal{C}|} \int_{\mathcal{C}'} \sum_{\bnu \in \Lambda} \e^{\im \bp \cdot \bnu} \Psi_m(\by_1,\bp) \, \overline{\Psi_j(\by_1+\bnu,\bp)}  \,\rd \by_1 \,\rd \bz \,\rd \bq\\[4pt]
        =& \frac{1}{(2\pi)^d} \frac{1}{|\mathcal{B}|} \int_{\RR^d} \int_{\RR^d} \e^{\im \bq \cdot \bz} \hat{N}(\bq) \sum_{m,n=1}^2 \sigma_{mn}(t,\bx,\bp)  \Psi_l(\bz,\bp+\bq_0) \,  \overline{\Psi_n(\bz,\bp)}  \\
        & \times \frac{1}{|\mathcal{C}|} \int_{\mathcal{C}'} \Psi_m(\by_1,\bp) \, \overline{\Psi_j(\by_1,\bp)}   \,\rd \by_1 \,\rd \bz \,\rd \bq\\[4pt]
         =& \frac{1}{(2\pi)^d} \frac{1}{|\mathcal{B}|} \int_{\RR^d} \int_{\RR^d} \e^{\im \bq \cdot \bz} \hat{N}(\bq) \sum_{m,n=1}^2 \sigma_{mn}(t,\bx,\bp) \delta_{mj} \Psi_l(\bz,\bp+\bq_0) \, \overline{\Psi_n(\bz,\bp)} \,\rd \bz \,\rd \bq
    \end{split}
\end{equation*}
where we apply the change of variable 
\begin{equation*}
    \by_1 \mapsto \bz-\by_1, \quad \by_2 \mapsto \bz-\by_2
\end{equation*}
in the third inequality above, use the identity \eqref{26} in the fifth equality as well as the periodicity of the Bloch eigenfunction
    $\overline{\Psi_j(\by_1+\bnu,\bp)} = \e^{-\im \bp \cdot \bnu} \overline{\Psi_j(\by_1, \bp)}$
    in the last second equality above.
\section{Explicit form of the random system away from  band crossings}
\label{appendix-C}
We now show how the system \eqref{sigma_tilde_random} reduces to \cite[Eq.~(3.15)]{BFPR1999JSP} whenever $\bp$ is away from band crossings. Recalling the discussion in Section \ref{sec:reduction}, we assume that the bands touch at finitely many crossing points $\bp_*$ with a structure given by \eqref{eq:lin_cross} for some integers $r > 0$. In this case, assuming $|\bp - \bp_*| \gg \ve^{1/r}$ as $\ve \downarrow 0$, we can replace the terms in \eqref{sigma_tilde_random} after taking the weak limit as $\ve \downarrow 0$:
%Whenever $\bp$ is away from the band crossings, we can replace, in the sense of weak limit as $\ve \downarrow 0$,
\begin{equation}
    \begin{split}
        &\e^{- \frac{\im}{\ve}[E_l(\bp) - E_n(\bp)]t} \rightarrow \delta(E_l(\bp) - E_n(\bp)) \rightarrow \delta_{ln} \, ,\\
        &\e^{-\frac{\im}{\ve}[E_m(\bq) - E_{m'}(\bq) + E_l(\bp) - E_j(\bp)] t} \rightarrow \delta(E_m(\bq) - E_{m'}(\bq) + E_l(\bp) - E_j(\bp)) \, , \\
        &\e^{-\frac{\im}{\ve}[E_{m'}(\bp) - E_{j}(\bp)] t} \rightarrow \delta( E_{m'}(\bp) - E_{j}(\bp) ) \rightarrow \delta_{m' j} \, .
    \end{split}
\end{equation}
and therefore we have: 
\begin{small}
    \begin{equation}\label{system_away}
    \begin{split}
    &\partial_t \widetilde{\sigma}_{jl}(t,\bx,\bp) + \nabla_\bx \widetilde{\sigma}_{jl}(t,\bx,\bp) \cdot  \left< (-\im \nabla_\bz) \Psi_{l} \,, \Psi_l \right>_{\mathcal{C}} \\
    =& \sum_{m = 1}^2 \frac{1}{(2\pi)^{d-1}} \frac{1}{|\mathcal{B}|} \sum_{\bmu \in \Lambda^*} \int_{\mathcal{B}}   \hat{R}(\bp-\bq+\bmu) \delta(E_l(\bp)-E_m(\bq)) \overline{A_{jm}(\bp, \bq-\bmu)}\\[6pt]
    & \times \Big[ \sum_{m' = 1}^2
    \delta(E_m(\bq) - E_{m'}(\bq) +  E_l(\bp) - E_j(\bp))
    \widetilde{\sigma}_{mm'}(\bq) A_{lm'}(\bp,\bq-\bmu) - 
    \widetilde{\sigma}_{jl}(\bp)  A_{jm}(\bp, \bq-\bmu) \Big] \,\rd \bq.
    \end{split}
\end{equation}
\end{small}
\eqref{system_away}  then provides the explicit dynamics for the diagonal terms $\widetilde{\sigma}_{jl}, j = l$ and off-diagonal terms $\widetilde{\sigma}_{jl}, j \neq l$, respectively:
\begin{itemize}
    \item When $j = l$, $\delta(E_m(\bq) - E_{m'}(\bq) +  E_l(\bp) - E_j(\bp))$ in \eqref{system_away} simplifies to
    \begin{equation*}
        \delta(E_m(\bq) - E_{m'}(\bq) +  E_l(\bp) - E_j(\bp)) \rightarrow \delta_{m m'}
    \end{equation*}
        so that we have, after replacing $\left<  (-\im \nabla_\bz)  \Psi_{j} \,, \Psi_j \right>_{\mathcal{C}}$ by $\nabla_{\bp} E_j(\bp)$,
    \begin{equation}
    \begin{split}
    &\partial_t \widetilde{\sigma}_{jj}(t,\bx,\bp) + \nabla_\bx \widetilde{\sigma}_{jj}(t,\bx,\bp) \cdot \nabla_{\bp} E_j(\bp)  \\
    =& \sum_{m = 1}^2 \frac{1}{(2\pi)^{d-1}} \frac{1}{|\mathcal{B}|} \sum_{\bmu \in \Lambda^*} \int_{\mathcal{B}}   \hat{R}(\bp-\bq+\bmu) \delta(E_l(\bp)-E_m(\bq)) |A_{jm}(\bp, \bq-\bmu)|^2 \\[6pt]
    & \times \Big[
    \widetilde{\sigma}_{mm}(\bq) - 
    \widetilde{\sigma}_{jj}(\bp)\Big] \,\rd \bq,
    \end{split}
\end{equation}
which is exactly \cite[Eq.~(3.15)]{BFPR1999JSP}.\\
    \item When $j \neq l$, a case-by-case analysis (listing all cases for $j,l,m,m' \in \{1,2\}$) shows that we can replace $\delta(E_m(\bq) - E_{m'}(\bq) + E_l(\bp) - E_j(\bp))$ by
    \begin{equation*}
    \delta(E_m(\bq) - E_{m'}(\bq) + E_l(\bp) - E_j(\bp)) \rightarrow \delta_{jm} \delta_{lm'} \delta(E_j(\bq) - E_{l}(\bq) +  E_l(\bp) - E_j(\bp)),
    \end{equation*}
    so that \eqref{system_away} becomes:
    \begin{small}
    \begin{equation}
    \begin{split}
    &\partial_t \widetilde{\sigma}_{jl}(t,\bx,\bp) + \nabla_\bx \widetilde{\sigma}_{jl}(t,\bx,\bp) \cdot  \left< (-\im \nabla_\bz) \Psi_{l} \,, \Psi_l \right>_{\mathcal{C}} \\
    =& \sum_{m = 1}^2 \frac{1}{(2\pi)^{d-1}} \frac{1}{|\mathcal{B}|} \sum_{\bmu \in \Lambda^*} \int_{\mathcal{B}}   \hat{R}(\bp-\bq+\bmu) \delta(E_l(\bp)-E_m(\bq)) \overline{A_{jm}(\bp, \bq-\bmu)}\\[6pt]
    & \times \Big[ 
    \delta(E_j(\bq) - E_{l}(\bq) +  E_l(\bp) - E_j(\bp))
    \widetilde{\sigma}_{jl}(\bq)   A_{ll}(\bp,\bq-\bmu) - 
    \widetilde{\sigma}_{jl}(\bp)   A_{jm}(\bp, \bq-\bmu) \Big] \,\rd \bq.
    \end{split}
    \end{equation}
    \end{small}
    In particular, each $\widetilde{\sigma}_{jl}$ evolves under a closed equation, ensuring that if the off-diagonal terms $\widetilde{\sigma}, j\neq l$ are initially zero, they remain zero for all $t$. This confirms that the off-diagonal $\sigma$ terms can be neglected away from band crossings as expected.
\end{itemize}

\section*{Acknowledgments}
We acknowledge helpful discussions with Guillaume Bal, John Schotland and Michael Hott. We also thank the anonymous reviewers for their valuable comments and suggestions, which significantly helped improve the quality of the paper.

\bibliographystyle{siamplain}
\bibliography{references}
\end{document}